\documentclass[12pt,twoside]{article}

\usepackage[
  left=3cm,right=2.5cm,top=3.1cm,bottom=2cm,headheight=2cm,headsep=1.2cm,footskip=1.4cm,twoside
]{geometry}
\usepackage[margin=1cm,font=small]{caption}
\usepackage{amsmath,amssymb}
\usepackage{lineno}
\usepackage{enumerate}
\usepackage{url}
\usepackage{subfigure}
\usepackage{pstricks}
\newcmykcolor{darkgreen}{1 0 0.6 0.5}  

\usepackage{epsfig}
\usepackage{epsf}
\usepackage{cite}   

\newlength{\nseparation}
\input{babarsymMerge-chka}

\usepackage{cite}   

\newcommand\Amptpbar{\kern 0.18em\overline{\kern -0.18em {\cal A}}_{3\pi}}

\newcommand\Amptpbarkappa{\kern 0.18em\overline{\kern -0.18em A}^{\kappa}{}}
\newcommand\Amptpbarsigma{\kern 0.18em\overline{\kern -0.18em A}^{\sigma}{}}

\newcommand\Nbpm{{\kern 0.18em\overline{\kern -0.18em N}}^{+-}}
\newcommand\Nbmp{{\kern 0.18em\overline{\kern -0.18em N}}^{-+}}

\newcommand\BRpmb{{\cal \kern 0.18em\overline{\kern -0.18em  B}}{}_{\rho\pi}^{+-}}
\newcommand\BRmpb{{\cal \kern 0.18em\overline{\kern -0.18em  B}}{}_{\rho\pi}^{-+}}

\newcommand\BRipmb{{\cal \kern 0.18em\overline{\kern -0.18em  B}}{}_{\rho^+\pi^-}}
\newcommand\BRimpb{{\cal \kern 0.18em\overline{\kern -0.18em  B}}{}_{\rho^-\pi^+}}

\newcommand\Abar{\kern 0.18em\overline{\kern -0.18em A}{}}

\def\journalL#1#2#3#4#5{\journal{#1 #2}{#3}{#4}{#5}}
\def\journal#1#2#3#4{#1~{\bf #2}, #3 (#4)}

\def\PRL#1#2#3{\journal{Phys.\ Rev. Lett.}{#1}{#2}{#3}}

\newcommand{\etal}{\emph{et al.}\xspace}

\begin{document}
\begin{titlepage}
  {\raggedleft CP3-12-01 \\
  }
  \vskip 2em
  {\centering
    \large 
    H.~Lacker$^{a}$, A.~Menzel$^{a}$, and F.~Spettel$^{a}$ [CKMfitter group],\\ 
    D. Hirschb\"uhl$^{b}$, J.~L\"uck$^{c}$, F.~Maltoni$^{d}$,  W.~Wagner$^{b}$, M.~Zaro$^{d}$\\
  }
  \vskip 2em
  {\centering {\Large Model-independent extraction of $|V_{tq}|$ matrix elements 
                      from top-quark measurements at hadron colliders.  }}
  \vskip 2.2em
  {
    \noindent
    Current methods to extract the quark-mixing matrix element $|V_{tb}|$ from 
    single-top production measurements assume that $|V_{tb}|\gg |V_{td}|, |V_{ts}|$: 
    top quarks decay into $b$ quarks with $100\%$ branching fraction, $s$-channel single-top 
    production is always accompanied by a $b$ quark and initial-state contributions from $d$ 
    and $s$ quarks in the $t$-channel production of single top quarks are neglected. 
    Triggered by a recent measurement of the ratio 
    $R=\frac{|V_{tb}|^{2}}{|V_{td}|^{2}+|V_{ts}|^{2}+|V_{tb}|^{2}}=0.90 \pm 0.04$ performed 
    by the D0 collaboration, we consider a $|V_{tb}|$ extraction method that takes into 
    account non zero $d$- and $s$-quark contributions both in production and decay.   
    We propose a strategy  that allows to extract consistently and in a model-independent 
    way the quark mixing matrix elements $|V_{td}|$, $|V_{ts}|$, and $|V_{tb}|$ from the 
    measurement of $R$ and from single-top measured event yields. As an illustration,  
    we apply our method to the Tevatron data using a CDF analysis of the measured 
    single-top event yield with two jets in the final state one of which is identified 
    as a $b$-quark jet.
    We constrain the $|V_{tq}|$ matrix elements within a four-generation scenario by 
    combining the results with those obtained from direct measurements in flavor 
    physics and determine the preferred range for the top-quark decay width within 
    different scenarios.
   }
    \vskip 1.5em \noindent
{\small \em $^{a}$ Humboldt-Universit\"at zu Berlin,
                   Institut f\"ur Physik,
                   Newtonstr. 15,
                   D-12489 Berlin, Germany,
                   {e-mail: lacker@physik.hu-berlin.de}}\\[0.2cm] 
{\small \em $^{b}$ Bergische Universit\"at Wuppertal,
                   Fachbereich C - Experimentelle Elementarteilchphysik,
                   Gau{\ss}str. 20,
                   D-42119 Wuppertal, Germany,}\\[0.2cm]
{\small \em $^{c}$ Karlsruhe Institute of Technology,
                   Institut f\"ur Experimentelle Kernteilchphysik,
                   Wolfgang-Gaede-Str. 1,
                   D-76131 Karlsruhe, Germany,}\\[0.2cm]
{\small \em $^{d}$ Centre for Cosmology, Particle Physics and Phenomenology (CP3),
                   Universit\'e catholique de Louvain,
                   Chemin du Cyclotron 2, 
                   B-1348 Louvain-la-Neuve, Belgium.}

\end{titlepage}


\noindent

\section{Introduction}

The top quark is not only the latest discovered and the heaviest particle in 
the Standard Model (SM), but also the only known fermion with a natural mass 
of order of the weak scale. For this reason in many SM extensions, from weakly 
interacting theories at the TeV scale, such as SUSY, to strongly interacting 
ones, it often plays a special role. This motivates the efforts aiming at 
measuring its properties with increasing accuracy and looking for significant 
deviations from theoretical predictions. 

At hadron colliders, the top quark is mainly produced in top-antitop pairs via 
strong interactions. 
However, the pure-electroweak production of a single top (or anti-top) quark 
has a remarkably competitive cross-section, and therefore can be very helpful 
in providing complementary information on top-quark properties. 
In the SM, the production of single top quarks occurs  via three different 
channels: the $s$- or $t$-channel exchange of a $W$ boson, and the associated 
$t W$ production. 
At the Tevatron, whose data we focus on in this paper, the $t W$ channel is 
negligible compared to the other two mechanisms, because of the smaller phase 
space available for the two heavy particles and the low gluon luminosity. 
In the $s$-channel, the top quark is produced from an intermediate $W^*$ boson 
in association with a light, down-type quark $q$, 
with rates proportional to the CKM elements $\left|V_{tq}\right|^2$~\cite{CKM}. 
In the $t$-channel the $W$ boson is exchanged between two quark lines allowing 
an initial state light quark $q$ to turn into the top quark. 
In this case the production rates are sensitive both to $\left|V_{tq}\right|^2$ 
and to the corresponding $q$ density inside the proton. In the Standard Model 
with three generations (3SM), $3 \times 3$ unitarity constrains the CKM element 
$V_{tb}$ to be very close to one ($|V_{tb}|=0.999146^{+0.000048}_{-0.000016}$~\cite{CKMfitter,CKMfitterSM}), 
and to overwhelm in size both $V_{td}$ and $V_{ts}$. Therefore contributions to 
the total $s$- and $t$-channel cross-sections involving light quarks other than 
the $b$ have usually been neglected in experimental analyses. For the same reason, 
the top-quark branching ratio into a $b$ quark and a $W$ boson
\begin{equation}
R=\frac{|V_{tb}|^{2}}{|V_{td}|^{2}+|V_{ts}|^{2}+|V_{tb}|^{2}}\,,
\end{equation}
is very close to one in the 3SM. As a result,  it is normally assumed to be equal 
to one in top-quark related analyses. However, it is clear that any analysis aiming 
at directly and jointly constraining $|V_{tb}|$  from single top production 
measurements and $R$, should not rely on the assumption 
$|V_{tb}|\gg |V_{td}|, |V_{ts}|$\cite{Alwall}.
For quite some time the most precise measurement of $R$ in $t \bar{t}$ production 
events using 0, 1 and 2 $b$-tagged jets came from D0~\cite{D0_Rmeasurement}:
$R=0.97^{+0.09}_{-0.08}$. In the 3SM this translates to the somewhat weak constraint 
$H = \sqrt{|V_{td}|^{2} + |V_{ts}|^{2}}/|V_{tb}| = \sqrt{(1-R)/R}< 0.49$ at $95\%$ 
confidence level (CL). Moreover, very recently, D0~\cite{D0_Rmeasurement_new} has 
presented  a much more precise measurement of $R$ giving $R=0.90 \pm 0.04$. 
The result points to a rather important deviation of $|V_{tb}|$ from one implying a 
$H \simeq 0.33$. This result renders the $|V_{tb}|\gg |V_{td}|, |V_{ts}|$ assumption 
untenable in any consistent extraction of $|V_{tb}|$ from both $R$ and single top 
production data.

As long as $R$ can significantly deviate from 1, one therefore needs to take into 
account contributions from $d$ and $s$ quarks in the production of single top 
quarks in the $t$-channel. For example, if a fourth generation of quarks (denoted 
$b'$ and $t'$) and leptons were realized in Nature  (4SM) the values of $|V_{td}|$ 
and $|V_{ts}|$ could be significantly larger than in the 3SM and the cross section 
in the $t$-channel could be modified by sizable $d$- and $s$-quark contributions, 
leading to detectable deviations from the 3SM, possibly showing up before a direct 
observation of the heavy quarks is possible.
In fact, direct limits on $V_{tq}$ CKM matrix elements provide very useful 
information on constraining and even excluding a 4SM. Extensive studies have 
recently appeared~\cite{Chanowitz,Lenz,London} which claim strong bounds on 
$|V_{tb}|$ using a combination of precision EW observables and flavor physics. 
Moreover, stronger and stronger bounds on the fourth generation quarks are 
quickly being set by the Tevatron and LHC 
experiments~\cite{lplus4jets,Collaboration:2012bt,Collaboration:2012xc,OppositeSignDileptons_ATLAS,OppositeSignDileptonsplusbjetsplusMissingET_CMS,Leptonsplusbjets_CMS,SameSignDileptonsplusjetplusbjetplusMissingET,bprimesearch_CDF,SameSignLeptons_ATLAS,SameSignLeptonsbjets_CMS,Inclusive_CMS}. 
It is interesting to note, however, that all these analyses contain 
assumptions on lifetimes and branching fractions, and therefore to some 
extent are model dependent. 
Designing a fully model- and assumption-independent analysis binding a fourth 
generation turns out to be not such an easy task and any complementary information 
that can be gathered is clearly welcome.

In this paper, we present the first quantitative analysis of single top production 
measurements which does not rely on the assumption $R=1$ and hence takes into 
account the effects from the CKM elements $|V_{td}|$ and $|V_{ts}|$ when extracting 
$|V_{tb}|$ in a truly model-independent way. The method refines and extends a 
simplified proposal first presented in Ref.~\cite{Alwall} and aims at extracting 
simultaneously constraints on the CKM elements $|V_{td}|$, $|V_{ts}|$ and $|V_{tb}|$ 
from $R$ and from the measured single top rates containing a $W$ boson and exactly 
two jets in the final state where either one or both jets are identified as a 
$b$-quark jet. Our extraction method is model-independent in the sense that it does 
not assume any hierarchy or flavor texture and can equally be applied to a 4SM 
scenario or to a model with a vector-like heavy quark. 
For the sake of illustration, we apply it to the the CDF published data with one 
identified $b$-quark jet out of two reconstructed jets. We then combine our results 
with direct measurements from flavor physics using the CKMfitter package~\cite{CKMfitter}  
and find the best value for the top quark decay width as well as constraints on the CKM elements 
in 3SM and 4SM scenarios.

The paper is organized as follows. We first outline the strategy followed in the 
analysis of single top production data. We then discuss in detail the sources 
of uncertainties and in particular those of systematic nature. Finally, we 
present the results of our simplified analysis based on the Tevatron data. 
We leave our conclusions and the outlook to the last section.

\section{Strategy}
\label{sec:strategy}

In this section we describe how an  analysis of the single top measurements 
at an hadron collider that is  free of assumptions on the CKM matrix elements 
could be performed. 
Our approach follows that proposed in Ref.~\cite{Alwall} and it could be used 
at the Tevatron as well as at the LHC. A consistent application requires direct 
access to the experimental data and a fully-fledged experimental analysis.  
As we will explain in the following, for the sake of illustration, we have 
only used published information from the Tevatron analyses on single top 
production in the two-jet final state and therefore only partially exploited 
the full potential of our approach. 

We express measured event yields in terms of:
\begin{enumerate}
\item the integrated luminosity $\mathcal{L}$,
\item the reconstruction efficiencies (which contain acceptance, trigger efficiencies, selection efficiencies including, for instance, $b$-tagging), 
\item total cross sections for the $s$- and 
$t$-channel~\cite{NLO_SingleTop_CrossSection_Harris,NLO_SingleTop_CrossSection_CampbellEllisTramontano,NLO_SingleTop_CrossSection_Kidonakis,Campbell:2009gj,Campbell:2009ss},
\item the CKM matrix elements $|V_{td}|$, 
$|V_{ts}|$ and $|V_{tb}|$,
\item the measured branching fraction $R$. 
\end{enumerate}

We denote the $s$-channel cross section as $\sigma^{s}$, and for the $t$-channel 
we distinguish between the cross sections induced by the initial quark flavour 
$q=d, s, b$ from which the top quark is produced: $\sigma_{d}^{t}$, $\sigma_{s}^{t}$, 
and $\sigma_{b}^{t}$.

Single top production is identified by selecting events with a lepton of high 
transverse momentum ($p_{T}$) indicating a $W$-boson decay and two or more 
reconstructed jets. In addition, one requires that at least one of these jets 
is tagged as a $b$-quark jet. The fully inclusive sample can be then organized 
in bins with a given jet multiplicity. 
The two-jet bin has  the highest sensitivity to single top as one expects only 
two jets in signal events with no extra radiation, $i.e.$ at the Born level, 
while more jets characterise the main $t \bar t$ background. However, in the 
current analyses, the three-jet (and even the four-jet) bin can provide additional
sensitivity to the signal when extra radiation is present and important information 
on the backgrounds.  
For the sake of illustration, we consider the number of single top signal events 
after background subtraction  (top and non top) classified by exactly one, 
respectively, two $b$-quark tagged jets in the two-jet final state. Extension 
to the three-jet final state can be done along the same lines and is 
straightforward. We note that in order to be consistent the effects of the general 
assumptions on the $|V_{tq}|$ CKM matrix elements have to be included also in the 
$t \bar t$ background. We therefore provide the corresponding rates for the top 
background at the end of each of the following subsections.

\subsection{Final state with one $b$-quark jet}

In the $s$ channel, the final state top-quark is accompanied by a light quark 
$q=d ,s, b$: $q + t$. The top quark decays subsequently into a light 
quark $q'=d, s, b$ plus a $W$ boson: $t \rightarrow q' + W$. We denote the 
final efficiencies to select such events as $\epsilon^{s}_{q (t \rightarrow q' W)}$. 
More in detail, the following efficiencies that depend both on production and 
decay of the top quark are considered:
\begin{itemize}
\item $\epsilon^{s}_{b (t \rightarrow b W)}$: production of $b + t$ with $t \rightarrow b W$,
\item $\epsilon^{s}_{b (t \rightarrow d W)}$: production of $b + t$ with $t \rightarrow d W$,
\item $\epsilon^{s}_{b (t \rightarrow s W)}$: production of $b + t$ with $t \rightarrow s W$, 
\item $\epsilon^{s}_{d (t \rightarrow b W)}$: production of $d + t$ with $t \rightarrow b W$,
\item $\epsilon^{s}_{d (t \rightarrow d W)}$: production of $d + t$ with $t \rightarrow d W$,
\item $\epsilon^{s}_{d (t \rightarrow s W)}$: production of $d + t$ with $t \rightarrow s W$, 
\item $\epsilon^{s}_{s (t \rightarrow b W)}$: production of $s + t$ with $t \rightarrow b W$, 
\item $\epsilon^{s}_{s (t \rightarrow d W)}$: production of $s + t$ with $t \rightarrow d W$,
\item $\epsilon^{s}_{s (t \rightarrow s W)}$: production of $s + t$ with $t \rightarrow s W$.
\end{itemize}
In the following, we assume that to a very good approximation 
$\epsilon^{s}_{d (t \rightarrow d W)} = \epsilon^{s}_{d (t \rightarrow s W)} = \epsilon^{s}_{s (t \rightarrow d W)} =\epsilon^{s}_{s (t \rightarrow s W)}$,
$\epsilon^{s}_{d (t \rightarrow b W)} = \epsilon^{s}_{s (t \rightarrow b W)}$,
and $\epsilon^{s}_{b (t \rightarrow d W)} = \epsilon^{s}_{b (t \rightarrow s W)}$.\\
Under these assumptions, for $s$-channel production the expected event yield for 
$W$ plus two jets in the final state where one jet is identified as a $b$-quark 
jet and the other one as a non-$b$-quark-jet is given by:
\begin{eqnarray}\label{Eq:1bjet_S_channel_1}
N_{1 b jet}^{2 jets,s} &=& \mathcal{L} \cdot \sigma^{s} \cdot R \nonumber \\ 
&[&|V_{tb}|^{2} \epsilon^{s}_{b (t \rightarrow b W)} \nonumber \\ 
&+& (|V_{td}|^{2}+|V_{ts}|^{2}) \nonumber \\
&\cdot&(\epsilon^{s}_{b (t \rightarrow d W)} + \epsilon^{s}_{d (t \rightarrow b W)} + \frac{1-R}{R} \epsilon^{s}_{d (t \rightarrow d W)})]
\end{eqnarray}
It is reasonable to assume (and easy to check) that compared to $\epsilon^{s}_{d (t \rightarrow b W)}$, $\epsilon^{s}_{b (t \rightarrow d W)}$, 
and $\epsilon^{s}_{b (t \rightarrow b W)}$ the efficiency $\epsilon^{s}_{d (t \rightarrow d W)}$
is small. In addition, this efficiency is multiplied by the factor $(1-R)/R$ which is at
most of order 0.1/0.9. We can therefore neglect the last term in Eq.~(\ref{Eq:1bjet_S_channel_1})
 and write
\begin{eqnarray}\label{Eq:1bjet_S_channel_2}
N_{1 b jet}^{2 jets,s} &=& \mathcal{L} \cdot \sigma^{s} \cdot R \nonumber \\ 
&[&|V_{tb}|^{2} \epsilon^{s}_{b (t \rightarrow b W)}
\nonumber \\
                 &+& (|V_{td}|^{2}+|V_{ts}|^{2}) (\epsilon^{s}_{b (t \rightarrow d W)} + \epsilon^{s}_{d (t \rightarrow b W)})]\,.
\end{eqnarray}
In general, one expects $\epsilon^{s}_{b (t \rightarrow d W)} \approx \epsilon^{s}_{d (t \rightarrow b W)}$.
However, in an analysis where one requires the invariant mass of the $\ell \nu b$
final state to lie in a window around the top quark mass one expects the efficiency
$\epsilon^{s}_{d (t \rightarrow b W)}$ to be significantly higher than 
$\epsilon^{s}_{b (t \rightarrow d W)}$. For an analysis where one has a high efficiency
to tag a $b$-quark jet one expects $\epsilon^{s}_{b (t \rightarrow b W)}$ to be small
since one has a high probability to find two $b$-jets.\\

In the $t$-channel the top quark is produced from a light quark $q =d, s, b$ inside 
the proton or antiproton ($q \rightarrow t$) and the top quark decays into a light 
quark $q'=d, s, b$ plus a $W$ boson: $t \rightarrow q' + X$. We denote the 
final efficiencies as $\epsilon^{t}_{q \rightarrow (t \rightarrow q' W)}$. As a result, 
depending on the top-quark production and its decay modes, the following efficiencies 
have to be considered:
\begin{itemize}
\item $\epsilon^{t}_{b \rightarrow (t \rightarrow b W)}$: $b \rightarrow t$ production with $t \rightarrow b W$, 
\item $\epsilon^{t}_{b \rightarrow (t \rightarrow d W)}$: $b \rightarrow t$ production with $t \rightarrow d W$, 
\item $\epsilon^{t}_{b \rightarrow (t \rightarrow s W)}$: $b \rightarrow t$ production with $t \rightarrow s W$, 
\item $\epsilon^{t}_{d \rightarrow (t \rightarrow b W)}$: $d \rightarrow t$ production with $t \rightarrow b W$,
\item $\epsilon^{t}_{d \rightarrow (t \rightarrow d W)}$: $d \rightarrow t$ production with $t \rightarrow d W$, 
\item $\epsilon^{t}_{d \rightarrow (t \rightarrow s W)}$: $d \rightarrow t$ production with $t \rightarrow s W$, 
\item $\epsilon^{t}_{s \rightarrow (t \rightarrow b W)}$: $s \rightarrow t$ production with $t \rightarrow b W$,
\item $\epsilon^{t}_{s \rightarrow (t \rightarrow d W)}$: $s \rightarrow t$ production with $t \rightarrow d W$,
\item $\epsilon^{t}_{s \rightarrow (t \rightarrow s W)}$: $s \rightarrow t$ production with $t \rightarrow s W$.
\end{itemize}
Assuming $\epsilon^{t}_{d \rightarrow (t \rightarrow d W)}=\epsilon^{t}_{d \rightarrow (t \rightarrow s W)}$,
$\epsilon^{t}_{s \rightarrow (t \rightarrow d W)} = \epsilon^{t}_{s \rightarrow (t \rightarrow s W)}$,
and $\epsilon^{t}_{b \rightarrow (t \rightarrow d W)} = \epsilon^{t}_{b \rightarrow (t \rightarrow s W)}$,
the $t$-channel production the expected event yield for 
$W$ plus two jets in the final state where one jet is identified as a $b$-quark 
jet and the other one as a non-$b$-quark-jet is given by:
\begin{eqnarray}\label{Eq:1bjet_T_channel}
N_{1 b jet}^{2 jets,t} = \mathcal{L} \cdot R \cdot &[& \sigma_{d}^{t} |V_{td}|^{2} (\epsilon^{t}_{d \rightarrow (t \rightarrow b W)}
                               +\epsilon^{t}_{d \rightarrow (t \rightarrow d W)}\frac{1-R}{R})\nonumber\\
                &+&              \sigma_{s}^{t} |V_{ts}|^{2} (\epsilon^{t}_{s \rightarrow (t \rightarrow b W)}
                               +\epsilon^{t}_{s \rightarrow (t \rightarrow d W)}\frac{1-R}{R})\nonumber\\
                &+&              \sigma_{b}^{t} (|V_{tb}|^{2} \epsilon^{t}_{b \rightarrow (t \rightarrow b W)} \nonumber \\
                              &+&(|V_{td}|^{2} + |V_{ts}|^{2}) \epsilon^{t}_{b \rightarrow (t \rightarrow d W)} )].
\end{eqnarray}
Once again, one can safely neglect the terms containing the factor $(1-R)/R$.
The prediction for the event yield from single top production with one $b$-quark 
jet tagged in the final state and the other jet not tagged as a $b$-quark is 
obtained by adding then Eqs.~(\ref{Eq:1bjet_S_channel_2}) and~(\ref{Eq:1bjet_T_channel}):
\begin{eqnarray}\label{Eq:1bjet}
N_{1 b jet}^{2 jets} = \mathcal{L} \cdot R \cdot &[& \sigma^{s} (|V_{tb}|^{2} \epsilon^{s}_{b (t \rightarrow b W)} \nonumber \\
              &+& (|V_{td}|^{2}+|V_{ts}|^{2}) (\epsilon^{s}_{b (t \rightarrow d W)} + \epsilon^{s}_{d (t \rightarrow b W)})) \nonumber\\
                    &+&              \sigma_{d}^{t} |V_{td}|^{2} \epsilon^{t}_{d \rightarrow (t \rightarrow b W)} \nonumber\\
                    &+&              \sigma_{s}^{t} |V_{ts}|^{2} \epsilon^{t}_{s \rightarrow (t \rightarrow b W)} \nonumber\\
                    &+&              \sigma_{b}^{t} (|V_{tb}|^{2} \epsilon^{t}_{b \rightarrow (t \rightarrow b W)} \nonumber \\
                              &+&(|V_{td}|^{2} + |V_{ts}|^{2}) \epsilon^{t}_{b \rightarrow (t \rightarrow d W)} )].
\end{eqnarray}
We remark that, following the definition above, the efficiencies take also into 
account higher order effects. For example, at NLO $t$-channel production can 
explicitly give rise to a ``spectator $b$" from the initial gluon splitting at high
transverse momentum and in the central region, which can contribute to the 
$b$-jet counting and therefore affect $\epsilon^t_{b \to (t \to bW)}$. 
Such contribution is present in an inclusive generation such as that from 
Pythia~\cite{Pythia6}  based on the $2\to 2$ leading order process.
It is generated via initial state radiation, an approximation that can  be very crude.
However, it is clear that this effect has a rather mild impact on $N^{2jets}_{1bjet}$ 
as it is an higher-order effect and in general three jets will then be present 
in the final state. A refined and more accurate analysis should determine the 
efficiencies by means of 4-flavor based calculations~\cite{Campbell:2009ss,Campbell:2009gj} 
where the kinematics of the spectator $b$'s are at NLO and/or obtained from 
MC@NLO~\cite{MC@NLO} and/or POWHEG~\cite{POWHEG} based simulations~\cite{Frixione:2005vw,Alioli:2009je}.

As mentioned above, the $t \bar t$ background will also be affected by the values 
of the $|V_{tq}|$ CKM matrix elements and therefore its subtraction has to be done 
consistently. Denoting by $\sigma^{tt}$ the total $t\bar t $ cross section and by 
${{\epsilon}}_{(t \rightarrow q W)(t \rightarrow q W)}$,
${{\epsilon}}_{(t \rightarrow b W)(t \rightarrow q W)}$,
${{\epsilon}}_{(t \rightarrow b W)(t \rightarrow b W)}$ the efficiencies for having a two-jet final 
state with one $b$-tag from a $t\bar t $ event with both, one, no top weak decay 
into a light quark-jet respectively, we have:
\begin{eqnarray}\label{Eq:tt1bjet}
N_{1 b jet}^{tt} = \mathcal{L} \cdot \sigma^{tt} \cdot &[& (1-R)^2 {{\epsilon}}_{(t \rightarrow q W)(t \rightarrow q W)}  
\nonumber \\
&+& 2 R (1-R) {{\epsilon}}_{(t \rightarrow b W)(t \rightarrow q W)}
\nonumber \\
&+&   R^2 {{\epsilon}}_{(t \rightarrow b W)(t \rightarrow b W)}].
\end{eqnarray}
The first term can be neglected as one expects both $(1-R)^2$ and  
${{\epsilon}}_{(t \rightarrow q W)(t \rightarrow q W)} $ to be very small, while the relative importance 
of the second term with respect to the third critically depends on the $b$-tagging 
efficiency.

\subsection{Final state with two $b$-quark jets}

The case of a $W$ boson and two jets where both jets have been identified as a 
$b$-quark jet  can be dealt with in a similar way. Substituting the efficiencies 
$\epsilon$ by corresponding efficiencies $\tilde{\epsilon}$ we obtain
\begin{eqnarray}\label{Eq:2bjets_1}
N_{2 b jets}^{2 jets} = \mathcal{L} \cdot R \cdot &[& \sigma^{s} (|V_{tb}|^{2} {\tilde{\epsilon}}^{s}_{b (t \rightarrow b W)} \nonumber \\
                             &+& (|V_{td}|^{2}+|V_{ts}|^{2}) ({\tilde{\epsilon}}^{s}_{b (t \rightarrow d W)}
                                           + {\tilde{\epsilon}}^{s}_{d (t \rightarrow b W)}))\nonumber\\
            &+&              \sigma_{d}^{t} |V_{td}|^{2} ({\tilde{\epsilon}}^{t}_{d \rightarrow (t \rightarrow b W)}
                          +{\tilde{\epsilon}}^{t}_{d \rightarrow (t \rightarrow d W)}\frac{1-R}{R})\nonumber\\
            &+&              \sigma_{s}^{t} |V_{ts}|^{2} ({\tilde{\epsilon}}^{t}_{s \rightarrow (t \rightarrow b W)}
                          +{\tilde{\epsilon}}^{t}_{s \rightarrow (t \rightarrow d W)}\frac{1-R}{R})\nonumber\\
            &+&              \sigma_{b}^{t} (|V_{tb}|^{2} {\tilde{\epsilon}}^{t}_{b \rightarrow (t \rightarrow b W)}\nonumber\\
                           &+&(|V_{td}|^{2} + |V_{ts}|^{2}) {\tilde{\epsilon}}^{t}_{b \rightarrow (t \rightarrow d W)} )].
\end{eqnarray}
The hierarchy in the efficiencies $\tilde{\epsilon}$ between different processes 
will in general differ from the one in the corresponding efficiencies $\epsilon$. 
For example, the efficiencies ${\tilde{\epsilon}}^{t}_{d \rightarrow (t \rightarrow d W)}$ and 
${\tilde{\epsilon}}^{t}_{s \rightarrow (t \rightarrow d W)}$ are supposed to be 
very small since no $b$-quark jet is produced in the final state. In addition, 
the ratio of the CKM matrix elements squared multiplying these efficiencies further
lower the importance of these terms, and therefore one would usually neglect them 
 further along in the analysis.
The $t \bar t$ background rate can be written in complete analogy to Eq.~(\ref{Eq:tt1bjet}) and reads
\begin{eqnarray}
\label{Eq:tt2bjet}
N_{2 b jet}^{tt} = \mathcal{L} \cdot \sigma^{tt} \cdot &[& (1-R)^2 {\tilde{\epsilon}}_{(t \rightarrow q W)(t \rightarrow q W)}  
\nonumber \\
&+& 2 R (1-R) {\tilde{\epsilon}}_{(t \rightarrow b W)(t \rightarrow q W)}
\nonumber \\
&+& R^2 {\tilde{\epsilon}}_{(t \rightarrow b W)(t \rightarrow b W)}]\,,
\end{eqnarray}
where one expects the last term to provide the bulk of the events.

\section{Inputs}\label{Sec:Inputs}

\subsection{Single top measurements and $R$ ratio}\label{Sec:Inputs_SingleTopandR}

For the branching fraction $R$ we use the best measured value from 
D0~\cite{D0_Rmeasurement_new}: $R=0.90 \pm 0.04$.
For the $s$ channel of single top (-anti-top) cross section we take the 
NLO+NNLL value from Ref.~\cite{NLO_SingleTop_CrossSection_Kidonakis} as
resummation effects are quite important in the $s$-channel, increasing 
the result of the NLO computation by the $10-15\%$.
The value is $\sigma^{s} = 1.074 \pm 0.030 \pm 0.045 ^{+0.001} _{-0.005}$ 
where we have chosen the top-quark mass of $m_{t}=(172.0 \pm 1.6)~{\rm GeV}$ 
from the Particle Data Group (PDG) edition 2010~\cite{PDG2010}.
The first uncertainty is due to the parton density function (PDF) uncertainty 
quoted in Ref.~\cite{NLO_SingleTop_CrossSection_Kidonakis}. The second one is 
due to the top-quark mass uncertainty of $\delta m_{t}=1.6~{\rm GeV}$~\cite{PDG2010}.
To obtain the corresponding uncertainty we interpolated the values for the total 
cross-sections quoted in Ref.~\cite{NLO_SingleTop_CrossSection_Kidonakis} and 
computed the variations for  mass uncertainity of $1.6~{\rm GeV}$. 
The third uncertainty takes into account contributions from higher orders in 
perturbation theory and is obtained from renormalization and factorization 
scale variations. The scale uncertainty is treated as a Rfit 
uncertainty~\cite{CKMfitter}, that is, the cross section is allowed to vary 
within this range without changing its $\chi^{2}$ contribution in the fit. 
The PDF and $m_{t}$ uncertainties are treated as statistical uncertainties, 
that is, they are assumed to follow a Gaussian likelihood.\\

Cross sections for single top (-anti-top) production in the $t$ channel are
calculated at NLO for a top-quark mass of $m_{t}=172~{\rm GeV}$. 
For the $t$-channel we choose not to use a resummed cross-section because resummation effects are smaller ($<5\%$)
and comparable with the uncertainty obtained from scale variations.
To obtain the NLO total cross-sections for $b$, $s$ and $d$ initiated single-top production,
we used a modified version of MCFM v5.8~\cite{MCFM}, where the
$b$-quark PDF in the calculation of the $t$-channel cross section can be replaced 
by a $d$-quark or an $s$-quark PDF. 
The cross section values for $t$ plus $\bar{t}$ production (which are equal at the 
Tevatron) and the assigned uncertainties are listed in Table~\ref{TABLE_XS}. For the 
PDFs we used the MSTW2008 PDF-sets~\cite{MSTW2008} from which the PDF uncertainty 
is estimated. The second uncertainty quantifies the effect from varying the renormalisation 
and factorisation scale between $m_{t}/2$ and $2 m_{t}$. We also quote the uncertainty 
from varying the $t$-quark mass by $1.6~{\rm GeV}$. The $m_{t}$ uncertainty is treated 
as fully correlated between all cross sections. For $\sigma_{b}^{t}$ we add also an 
uncertainty coming from the $b$-quark mass~\cite{Campbell:2009gj}.

\begin{table}[Htp]
\renewcommand{\arraystretch}{1.3}
\centering
\begin{tabular}{|c|c|c|c|c|c|}\hline
Cross section    & value & PDF unc. & $m_{t}$ unc. & $m_{b}$ unc. & scale unc. \\
                 & [pb]  &   [pb]   &     [pb]     & [pb]         &  [pb] \\
\hline
$\sigma_{d}^{t}$ & $24.96$    & $\pm 0.84$    & $^{+0.49}_{-0.75}$ & -                 & $\pm 0.75$      \\
$\sigma_{s}^{t}$ & $ 6.48$    & $\pm 0.31$    & $^{+0.11}_{-0.16}$ & -                 & $\pm 0.20$      \\
$\sigma_{b}^{t}$ & $ 2.01$    & $\pm 0.07$    & $^{+0.06}_{-0.05}$ & $\pm 0.05$        & $\pm 0.06$      \\
\hline
\end{tabular}
\caption[TABLE_XS]
{NLO cross section for $t$-channel single top plus single antitop production at the 
 Tevatron for a center-of-mass energy $\sqrt{s}=1.96~{\rm TeV}$ calculated with 
 MCFM~\cite{MCFM} using the PDF sets taken from MSTW2008~\cite{MSTW2008}.
 The scale uncertainty is treated as a Rfit uncertainty~\cite{CKMfitter}, that is, 
 the cross section is allowed to vary within this range without changing its $\chi^{2}$ 
 contribution in the fit. The PDF, $m_{t}$ and $m_b$ uncertainties are treated as 
 statistical uncertainties, that is, they are assumed to follow a Gaussian likelihood.
}
\label{TABLE_XS}
\end{table}

For completeness we have also studied the impact of the correlations due to PDF uncertainities and scale variations
between the individual cross section calculations in the $t$ channel. To compute the correlation coefficients (or correlation cosines) for two set of numbers $\{X_i\}$ and and $\{Y_i\}$ corresponding to
the values of two cross-sections with varied scales, we used the formula:
\begin{equation}
    \rho=\frac{1}{4 \Delta X\Delta Y}\sum_i \left(X_i-X_{ref} \right)\left(Y_i-Y_{ref} \right)\,
\end{equation}
where it is understood that $X_i$ and $Y_i$ are computed with the same scales, 
\begin{eqnarray}
    \Delta X =\sqrt{\frac{1}{N}\sum_i\left(X_i -X_{ref}\right)^2}\,,\nonumber\\
    \Delta Y =\sqrt{\frac{1}{N}\sum_i\left(Y_i -Y_{ref}\right)^2}\,,
\end{eqnarray}
and $X_{ref}$, $Y_{ref}$ are the values computed with the reference renormalization and factorization scales.
For the correlations coming from PDF errors we used the MSTW prescription as given in expression in Eq.~(50) of Ref.~\cite{MSTW2008}.
The correlation coefficients between the $t$-channel cross sections due to the PDF uncertainty are found to be 
$\rho_{PDF}(\sigma_{d}^{t},\sigma_{s}^{t})=+0.067$, $\rho_{PDF}(\sigma_{d}^{t},\sigma_{b}^{t})=-0.273$,
and $\rho_{PDF}(\sigma_{s}^{t},\sigma_{b}^{t})=+0.111$. The correlations from the scale 
uncertainty are determined as $\rho_{scale}(\sigma_{d}^{t},\sigma_{s}^{t})=+0.809$, 
$\rho_{scale}(\sigma_{d}^{t},\sigma_{b}^{t})=-0.568$, and $\rho_{scale}(\sigma_{s}^{t},\sigma_{b}^{t})=-0.566$.
Within the set of inputs used we find that the quantitative results of 
our analysis (Sec.~\ref{NumericalResults}) do not exhibit a large effect 
from these correlations. \\

In our simplified study, we use results from CDF for a final state with two jets
one of which is identified as a $b$-quark jet~\cite{CDF_SingleTop_1,CDF_SingleTop_2} 
based on an integrated luminosity of $\mathcal{L}=(3.20 \pm 0.16)~{\rm fb^{-1}}$. 
The largest part of the observed candidate events was selected online with single
lepton triggers, electron or muon, where the charged leptons are detected in the
central part of the CDF-II dedector, that is equipped with fast track-finding.
In addition, electrons in the endcap calorimeters are used to trigger candidate 
events. The acceptance for these two classes of events defines the ``trigger lepton
coverage'' (TLC). In addition, there are also events that were recorded with a
missing transverse energy plus jet trigger where a muon is reconstructed offline. 
The acceptance of these events is named ``extended muon coverage'' (EMC).

Assuming $R=1$\footnote{The cross section extraction in the CDF analysis assumes 
also $R=1$ for the $t \bar{t}$ background estimate. While we cannot take this 
effect into account in our simplified analysis, it should be done in a complete one, as outlined in Sec.~\ref{sec:strategy}.}
the result for the sum of the measured $s$- and $t$-channel cross section reads 
$\sigma_{EMC}=(2.3^{+1.4}_{-1.1})~{\rm pb}$ and 
$\sigma_{TLC}=(1.7^{+0.7}_{-0.6})~{\rm pb}$. For the further analysis we 
symmetrise the uncertainties on these measured cross sections and calculate a 
weighted average of both results, $\sigma_{meas}=(1.83 \pm 0.58)~{\rm pb}$, which 
can be translated into a signal yield of $N_{1 b jet}^{2 jets} = 84.3 \pm 26.7$.
This signal yield is below the expected number of events of about 142.8 in the
3SM scenario which is obtained by setting $|V_{tb}|$, $|V_{ts}|$, and $|V_{td}|$
to their very well-known 3SM values.

The relevant efficiencies as quoted in Table~\ref{TABLE_EFF} have been obtained 
by using Ref.~\cite{CDF_SingleTop_2} accompanied by an additional study in 
which it was determined in what fractions the $b$-tagged jet comes from the 
top-quark decay, from the 2nd $b$ quark (in case of the $t$ channel) and from 
the top-production vertex ($s$ channel). Since we have not used a dedicated
simulation studying $t$-channel production from $d$- and $s$-quarks we set for
simplicity $\epsilon^{t}_{d \rightarrow (t \rightarrow b W)} = \epsilon^{t}_{s \rightarrow (t \rightarrow b W)}$.
One should note, however, that some difference between these efficiencies is 
expected as $s$-quark $t$-channel production has contributions only from sea quarks 
while $d$-quark $t$-channel production has contributions not only from sea quarks 
but also from valence quarks. As a consequence, the kinematic distributions for 
both event classes will be different.

We add an additional uncertainty of $10\%$ reflecting systematic uncertainties 
that are taken to be fully correlated between all effciencies.

\begin{table}[Htp]
\renewcommand{\arraystretch}{1.3}
\centering
\begin{tabular}{|c|c|c|c|}\hline
Efficiency                                        &  value       &  stat. unc.     & sys. unc.     \\
\hline
$\epsilon^{s}_{b (t \rightarrow b W)}$              &  $0.01601$   &  $\pm 0.00031$  &  $\pm 0.0016$ \\
$\epsilon^{s}_{b (t \rightarrow d W)}$              &  $0.01278$   &  $\pm 0.00024$  &  $\pm 0.0013$ \\
$\epsilon^{s}_{d (t \rightarrow b W)}$              &  $0.01278$   &  $\pm 0.00024$  &  $\pm 0.0013$ \\
$\epsilon^{t}_{b \rightarrow (t \rightarrow b W)}$  &  $0.01346$   &  $\pm 0.00054$  &  $\pm 0.0013$ \\
$\epsilon^{t}_{b \rightarrow (t \rightarrow d W)}$  &  $0.00062$   &  $\pm 0.00003$  &  $\pm 0.00006$ \\
$\epsilon^{t}_{d \rightarrow (t \rightarrow b W)}$  &  $0.01322$   &  $\pm 0.00052$  &  $\pm 0.0013$ \\
\hline
\end{tabular}
\caption[TABLE_EFF]
{Values and uncertainties used in this analysis for the efficiencies as defined 
 in the text. The first uncertainty is the statistical uncertainty from the 
 limited statistics of the Monte-Carlo simulation. The second uncertainty reflects 
 a global systematic uncertainty on the efficiency of order $10~\%$.
}
\label{TABLE_EFF}
\end{table}

\subsection{Extraction of $|V_{tb}|$ in a fourth generation scenario: additional constraints}
\label{Sec:4SMextraction}

The method outlined in the previous section allows to extract information on the 
CKM $|V_{tq}|$ matrix elements in a model-independent way.
In a specific scenario, however,  other constraints can be added by using other 
available information. Let us consider the case of a fourth generation and the data 
from flavor physics. In order to be free of any assumption,  we consider only constraints 
on CKM matrix elements $|V_{ij}|$  that are extracted from tree-level decay processes. 
These are $|V_{ud}|$, $|V_{us}|$, 
$|V_{ub}|$, $|V_{cd}|$, and $|V_{cb}|$ as quoted in Table~\ref{TABLE_CKM}. 
In addition, we use the measured branching fractions of leptonic $W$ decays 
$W \rightarrow \ell {{\nu}}_{\ell}$. 

\begin{table}[Htp]
\renewcommand{\arraystretch}{1.3}
\centering
\begin{tabular}{|c|c|c|c|c|}\hline
Input                                                       & value     & stat. unc.               & theo. unc.    & Ref.      \\
\hline
$|V_{ud}|$                                                  & $0.97413$ & $^{+0.00033}_{-0.00023}$ & -              & see text       \\
$|V_{us}|$                                                  & $0.2254$  & $\pm 0.0013$             & -              & \cite{FlaviaNet2010}\\
$|V_{ub}| \times 10^{-3}$                                   & $3.92$    & $\pm 0.09$               & $\pm 0.45$     & \cite{CKMfitterNP} \\
$|V_{cd}|$                                                  & $0.230$   & $\pm 0.011$              & -              & \cite{PDG2010} \\
$|V_{cb}| \times 10^{-3}$                                   & $40.89$   & $\pm 0.38$               & $\pm 0.59$     & \cite{CKMfitterNP}\\
$\mathcal{B}(W \rightarrow e {{\nu}}_{e})$       & $0.1075$  & $\pm 0.0013$             & -              & \cite{PDG2010} \\
$\mathcal{B}(W \rightarrow \mu {{\nu}}_{\mu})$   & $0.1057$  & $\pm 0.0015$             & -              & \cite{PDG2010} \\
$\mathcal{B}(W \rightarrow \tau {{\nu}}_{\tau})$ & $0.1125$  & $\pm 0.0020$             & -              & \cite{PDG2010} \\
\hline
\end{tabular}
\caption[TABLE_CKM]
{Values and uncertainties used in the analysis when adding aditional constraints
 on the $4 \times 4$ CKM matrix as described in the text. The first uncertainty 
 is a statistical uncertainty, the second uncertainty reflects a theoretical 
 uncertainty which is treated as an Rfit error as described in Ref.~\cite{CKMfitter}.
}
\label{TABLE_CKM}
\end{table}

As discussed in Ref.~\cite{LackerMenzel} the extraction  of $G_F$ within a fourth
generation scenario has an impact on the determination of the CKM matrix elements
from leptonic and semileptonic meson decays. In practice, however, only the 
extraction of $|V_{ud}|$ from super-allowed $\beta$ decays~\cite{TownerHardy} 
is visibly affected. Hence, for $|V_{ud}|$ we can simply take the value and uncertainty 
from Ref.~\cite{LackerMenzel}. For $|V_{us}|$ we adopt the average value obtained
from semileptonic kaon decays provided by Flavianet~\cite{FlaviaNet2010}. 
For $|V_{ub}|$, we take the average value quoted in Ref.~\cite{CKMfitterNP} 
based on the average values for $|V_{ub}|$ extracted by the Heavy Flavour 
Averaging Group (HFAG)~\cite{HFAG} from inclusive and exclusive charmless 
semileptonic $B$-meson decays. For $|V_{cd}|$, we use the value quoted 
by the PDG~\cite{PDG2010} based on deep-inelastic scattering
of neutrino on nucleons. 
For the matrix element $|V_{cb}|$, we take the average value quoted in 
Ref.~\cite{CKMfitterNP} based on the HFAG average values for $|V_{cb}|$~\cite{HFAG} 
from inclusive and exclusive charmed semileptonic $B$-meson decays.

The branching fraction for $W \rightarrow \ell {{\nu}}_{\ell}$, where $\ell$ 
is either an electron, muon or $\tau$, is predicted to be
\begin{eqnarray}\label{Eq:LeptonicWdecays}
\mathcal{B}(W \rightarrow \ell {{\nu}}_{\ell})=\frac{1}{3+3 \sum\limits_{i=u,c}\sum\limits_{j=d,s,b} |V_{ij}|^{2}(1+\frac{\alpha_s(m_W)}{\pi})}
\end{eqnarray}
In this prediction, one assumes $3 \times 3$ unitarity of the neutrino-mixing matrix.
In a 4SM scenario, the deviations from this assumption are much smaller than the
relevant experimental uncertainty when combined with other observables that constrain
the 4SM neutrino-mixing matrix elements, and, hence, can be neglected for the numerical
studies in the quark sector~\cite{LackerMenzel}.
The measured leptonic branching fractions of $W$-bosons provides a $3 \times 3$ 
unitarity test for the sum $\sum_{i=u,c}\sum_{j=d,s,b} |V_{ij}|^{2}$. In a 4SM 
scenario this measurement can be used to constrain $|V_{cs}|$ when using the 
precisely measured CKM elements $|V_{ud}|$, $|V_{us}|$, $|V_{ub}|$, $|V_{cd}|$, 
and $|V_{cb}|$. The resulting constraint, $|V_{cs}|=0.9733^{+0.00054}_{-0.00948}$
(at $1~\sigma$ level), is tighter than that one from semileptonic decays of $D_{s}$
mesons, $i.e.$ $|V_{cs}|=0.98 \pm 0.01 \pm 0.10$~\cite{PDG2010}, 
and also the one from leptonic $D_{s}$-decays, $|V_{cs}|=1.030 \pm 0.038$~\cite{PDG2010},
which prefers values above the unitarity bound.

For $\mathcal{B}(W \rightarrow \ell {{\nu}}_{\ell})$ we use the LEP 
averages quoted in Ref.~\cite{PDG2010}. The correlations between the measured 
branching fractions for the electron, muon, and $\tau$ final states are taken 
into account. Their values have been taken from Ref.~\cite{LEP} and refer to 
preliminary results which have not been published yet.  We assume that the final
results will not be very different.  To conclude, we note that the three 
measured leptonic $W$ branching fractions are not in perfect agreement. This 
leads to a rather large $\chi^{2}$ value in the numerical analysis but does not 
change any of the conclusions.

\section{Numerical Results}\label{NumericalResults}

We now discuss the results of our simplified analysis. To be concrete and study the impact
of different assumptions, we consider several methods which can be classified in three broad 
categories:
\begin{itemize}
\item We first consider $|V_{tb}|\gg|V_{td}|, |V_{ts}|$. As experimental information 
      we use only $N_{1 b jet}^{2 jets}$ and in Eq.~(\ref{Eq:1bjet}) we set $R$ 
      identically to one and hence $|V_{td}|=|V_{ts}|=0$. This is the method that has been 
      usually used by the Tevatron experiments to translate the measured cross section into 
      a constraint on $|V_{tb}|$. We call this the `$R=1$ method'.
\item In the general method, we use Eq.~(\ref{Eq:1bjet}) and extract from data information 
      on $N_{1 b jet}^{2 jets}$ and $R$. We leave the possibility to performing the analysis 
      either in the 3SM or in the 4SM. Accordingly, we call these extraction methods the 
      `3SM method' and `4SM method'. The `4SM method' is essentially equivalent to having 
      $|V_{tb}|$, $|V_{td}|$, and $|V_{ts}|$ determined without applying any unitarity 
      constraints as long as $|V_{td}|^2 + |V_{ts}|^2 + |V_{tb}|^2 \le 1$.\\
      On top of the `4SM method', one can add constraints on CKM elements
      from tree-level direct measurements as discussed in Sec.~\ref{Sec:4SMextraction}
      which then set additional constraints on $|V_{td}|$ and $|V_{ts}|$ thanks
      to $4 \times 4$ unitarity. We call this extraction method the `4SMTL method'.\\
\item
      The most general method which we apply is to consider $|V_{td}|$, $|V_{ts}|$, 
      and $|V_{tb}|$ as free parameters without applying any unitarity constraints. 
      This method leads to the same constraints as the `4SM method' as long as 
      $|V_{td}|^2 + |V_{ts}|^2 + |V_{tb}|^2 \le 1$ but also allows values larger than one 
      for these individual parameters. Such a fit model, which we call the `free CKM method', 
      allows to quantify non-Standard-Model couplings of the top quark to the light quark 
      flavours $d$, $s$, and $b$.\\
\end{itemize}
We note that in a recent analysis of the D0 collaboration, an approach similar to the `free CKM method' above has been used to 
      extract the top-quark decay width from the measurement of $R$ and the $t$-channel 
      cross section~\cite{D0TopWidth}. However, in the D0 analysis it is assumed that 
      one can write for the top-quark decay rate to $b$ quarks:
      \begin{eqnarray}\label{Eq:D0Method}
        \Gamma(t \rightarrow bW) = \Gamma(t \rightarrow bW)_{SM} \frac{\sigma^{t}}{\sigma^{t}_{SM}},
      \end{eqnarray}
      where $\Gamma(t \rightarrow bW)_{SM}$ is the 3SM prediction of the top-quark decay 
      rate to $b$ quarks, $\sigma^{t}$ the measured $t$-channel cross section, and 
      $\sigma^{t}_{(SM)}$ its 3SM prediction, that is, $t-s$ and $t-d$ couplings 
      are neglected in the $t$-channel cross section.

\subsection{Implications of the $R$ measurement for a 4SM scenario}

\begin{figure}
\resizebox{0.99\textwidth}{!}{
  \includegraphics{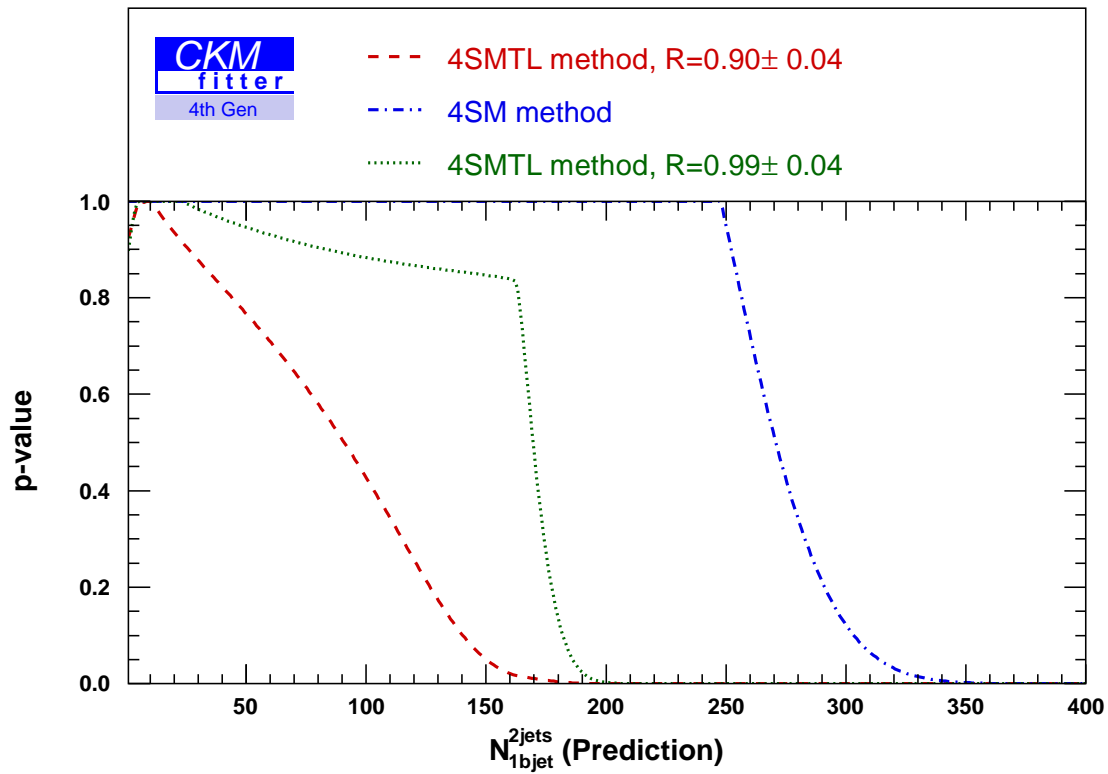}
}
\caption{Prediction of the event yield $N_{1 b jet}^{2 jets}$ as discussed in the text. 
         Blue dotted-dashed line: `4SM method'. 
         Red dashed line: `4SMTL method', $i.e.$ applying the tree-level constraints 
         on $|V_{ud}|$, $|V_{us}|$, $|V_{ub}|$, $|V_{cd}|$, 
         $|V_{cb}|$, and ${\cal{B}}(W \rightarrow \ell \nu_{\ell})$ in a the 4SM
         scenario as explained in the text. 
         Green-dotted line: `4SMTL method' assuming a different $R$ measurement 
         ($R=0.99 \pm 0.04$).}
\label{Yield_pred}       
\end{figure}

The D0 measurement of $R=0.90 \pm 0.04$ deviates by $2.5\sigma$ from the 3SM
expectation. In a 4SM scenario this measurement can be easily accommodated. 
Using the direct information on $|V_{ud}|$, $|V_{us}|$, $|V_{ub}|$, 
$|V_{cd}|$, $|V_{cb}|$, the matrix elements $|V_{td}|$ and $|V_{ts}|$ can be 
well constrained using $4 \times 4$ unitarity. When combining these inputs 
with $R=0.90 \pm 0.04$ one finds $|V_{td}| < 0.08$ and $|V_{ts}| < 0.31$ at 
$95.54\%$ Confidence Level (CL).
As a consequence, given the smallness of $|V_{td}|$ and $|V_{ts}|$, an $R$ value 
as small as $0.9$ can only be obtained if $|V_{tb}|$ is small as well. With our 
inputs the preferred value for $|V_{tb}|$ is found to be $0.21$ with upper limits 
of $0.78$ at $68.3~\%$ CL and $0.92$ at $95.54~\%$ CL.
This would lead to significantly smaller event yields in single top production
both at the Tevatron and at the LHC because all $s$- and $t$-channel contributions
would be CKM-suppressed. While the CDF result used in our analysis has a yield
below the 3SM expectation the single top production cross section value measured 
by D0 is larger than the 3SM prediction~\cite{Abazov:2011pt}. Moreover, the first 
single top production measurements at LHC point to cross sections as large as or 
even larger than the 3SM prediction~\cite{ATLAS_SingleTop,CMS_SingleTop}. 
Keeping in mind that the uncertainties are still large these measurements are hence 
not easily accommodated with the D0 $R$ measurement in a fourth generation scenario. 
This is illustrated (see Fig.~\ref{Yield_pred}) by predicting the single top yield 
$N_{1 b jet}^{2 jets}$ using the luminosity, cross sections and efficiencies from 
Sec.~\ref{Sec:Inputs} in the `4SM method', that is, without any additional constraints 
on the $4 \times 4$ CKM matrix.
The corresponding result (blue dotted-dashed line) shows that the yield could vary 
a lot with respect to the 3SM value of 142.8 events. However, when taking as inputs 
the constraints from $|V_{ud}|$, $|V_{us}|$, $|V_{ub}|$, $|V_{cd}|$, 
$|V_{cb}|$, and on the leptonic branching fractions of $W$-bosons, the expected yield 
(red dashed line in Fig.~\ref{Yield_pred}) clearly prefers smaller values although 
the 3SM value of 142.8 is still allowed. 

The single top cross section measured by CDF gives a measured event yield of 
$N_{1 b jet}^{2 jets}=84.3 \pm 26.8$, in reasonable agreement with the prediction 
in the `4SMTL method'.
Please note that a larger central value for $|V_{cs}|$ as favoured by leptonic 
$D_{s}$ decays would favour small $N_{1 b jet}^{2 jets}$ values even stronger 
since in this case the upper limit on $|V_{ts}|$ would shift towards smaller values.

For a comparison we also present the predicted event yield in the `4SMTL method'
(Fig.~\ref{Yield_pred}, green dotted line) under the assumption of a hypothetical 
$R$ value much closer to one: $R=0.99 \pm 0.04$. In this case, event yields as large 
or even slightly larger than the 3SM expectation are possible in the `4SMTL method'.
In other words, the small single top yield of the CDF analysis and the 
$R=0.90 \pm 0.04$ result of D0 can be accommodated within the `4SMTL method' while 
the other, large single top cross section results are difficult to explain in a 
`4SMTL method' with a $R$ value significantly smaller than one.

\subsection{Numerical results with the current inputs}

\begin{figure}
\resizebox{0.99\textwidth}{!}{
  \includegraphics{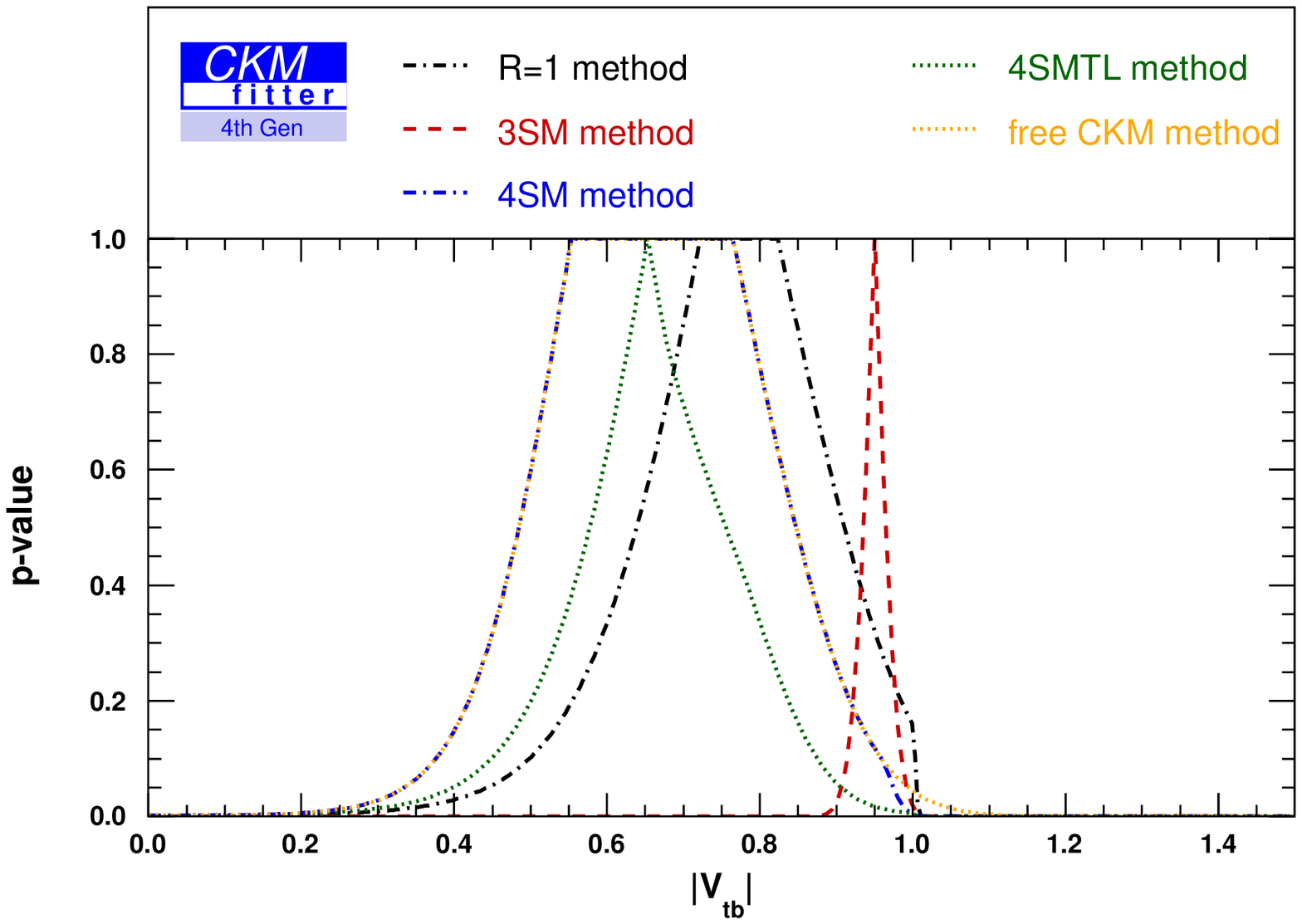}
}
\caption{Constraint on $|V_{tb}|$ from the single top analysis within 
         different scenarios using $N_{1 b jet}^{2 jets}=84.3 \pm 26.8$:
         black dashed-dotted curve: p-value obtained in the `$R=1$ method';
         red dashed curve: p-value obtained in the `3SM method' setting $R=0.90 \pm 0.04$;
         blue dashed-dotted curve: p-value obtained in the `4SM method' setting $R=0.90 \pm 0.04$;
         green dotted curve: p-value obtained in the `4SMTL method' setting $R=0.90 \pm 0.04$
         and using in addition constraints on $|V_{ud}|$, $|V_{us}|$, $|V_{ub}|$, 
         $|V_{cd}|$, 
         $|V_{cb}|$, and ${\cal{B}}(W \rightarrow \ell \nu_{\ell})$;
         orange dotted curve: p-calue obtained in the `free CKM method' setting $R=0.90 \pm 0.04$.}
\label{Vtb_normal}       
\end{figure}

With the numerical inputs from Sec.~\ref{Sec:Inputs} we obtain constraints on 
$|V_{tb}|$ as shown in Fig.~\ref{Vtb_normal}. The black dashed-dotted curve shows 
the p-value as a function of $|V_{tb}|$ for the `$R=1$ method' that neglects 
$d$- and $s$-quark contributions in single top production and top-quark decay.\\
The red dashed curve shows the constraint from the `3SM method' using the measured 
value of $R$.
Compared to the `$R=1$ method' the constraint on $|V_{tb}|$ is much stronger 
in the `3SM method'. This might look odd at first glance because $R$ carries 
now an experimental uncertainty. However, in the `3SM method', not only the 
single top yield but also $R$ itself provides a constraint on $|V_{tb}|$ thanks 
to $3 \times 3$ unitarity of the CKM matrix. The experimental information from 
$R$ is much more constraining than the measured value for the single top yield 
used in our analysis. The preferred value is $|V_{tb}|=0.95$ and $|V_{tb}|=1$
is disfavoured as a result of the $2.5~\sigma$ deviation of $R$ from $1$.
The blue dashed-dotted curve shows the constraint within the `4SM method' which 
turns out to be much less constraining than the one obtained in the `$R=1$ method'.

\begin{figure}
\resizebox{0.99\textwidth}{!}{
  \includegraphics{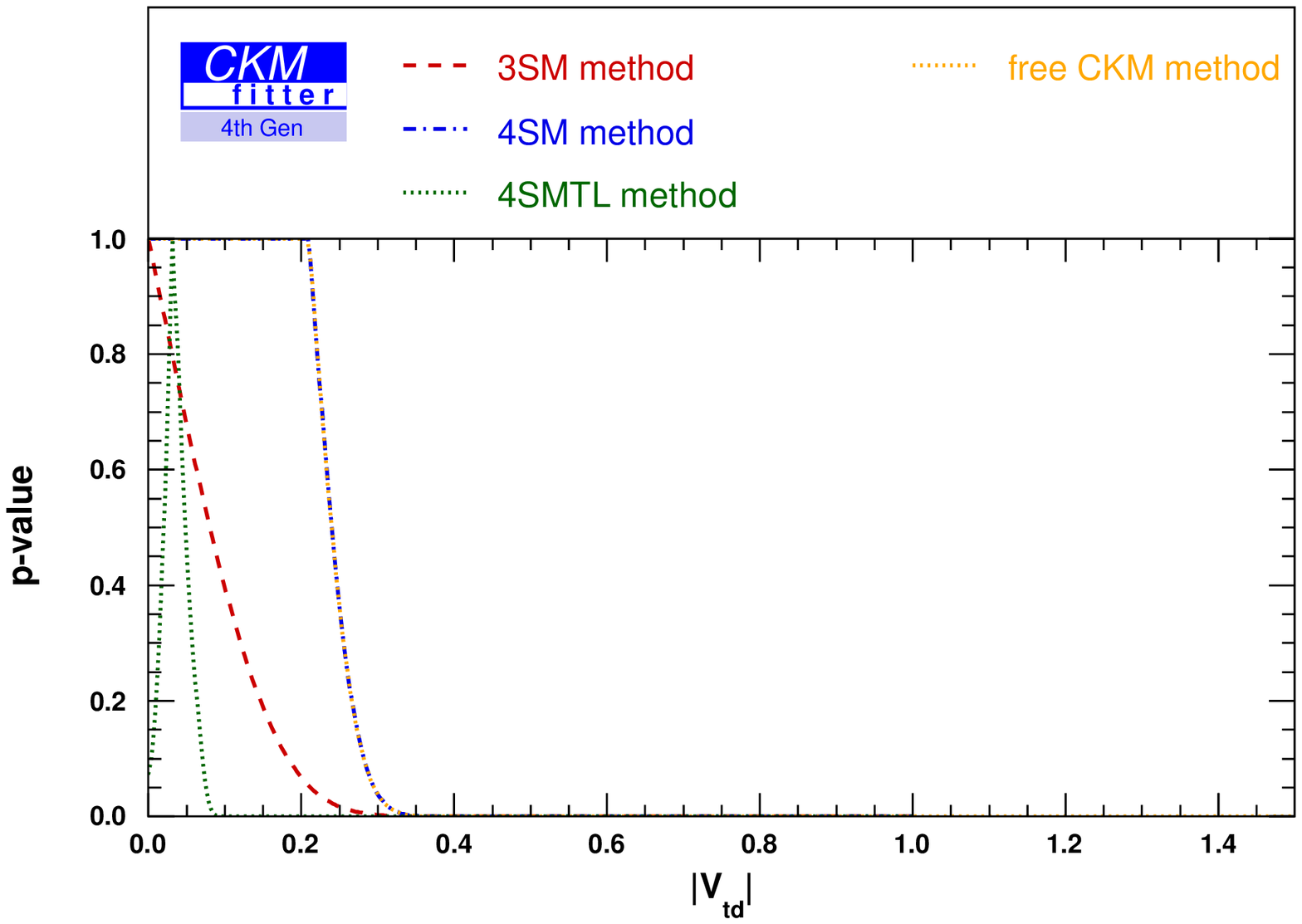}
}
\caption{Constraint on $|V_{td}|$ from the single top analysis within 
         different scenarios using $N_{1 b jet}^{2 jets}=84.3 \pm 26.8$.
         For the different colour-coding we refer to the caption of Fig.~\ref{Vtb_normal}.}
\label{Vtd_normal}       
\end{figure}

\begin{figure}
\resizebox{0.99\textwidth}{!}{
  \includegraphics{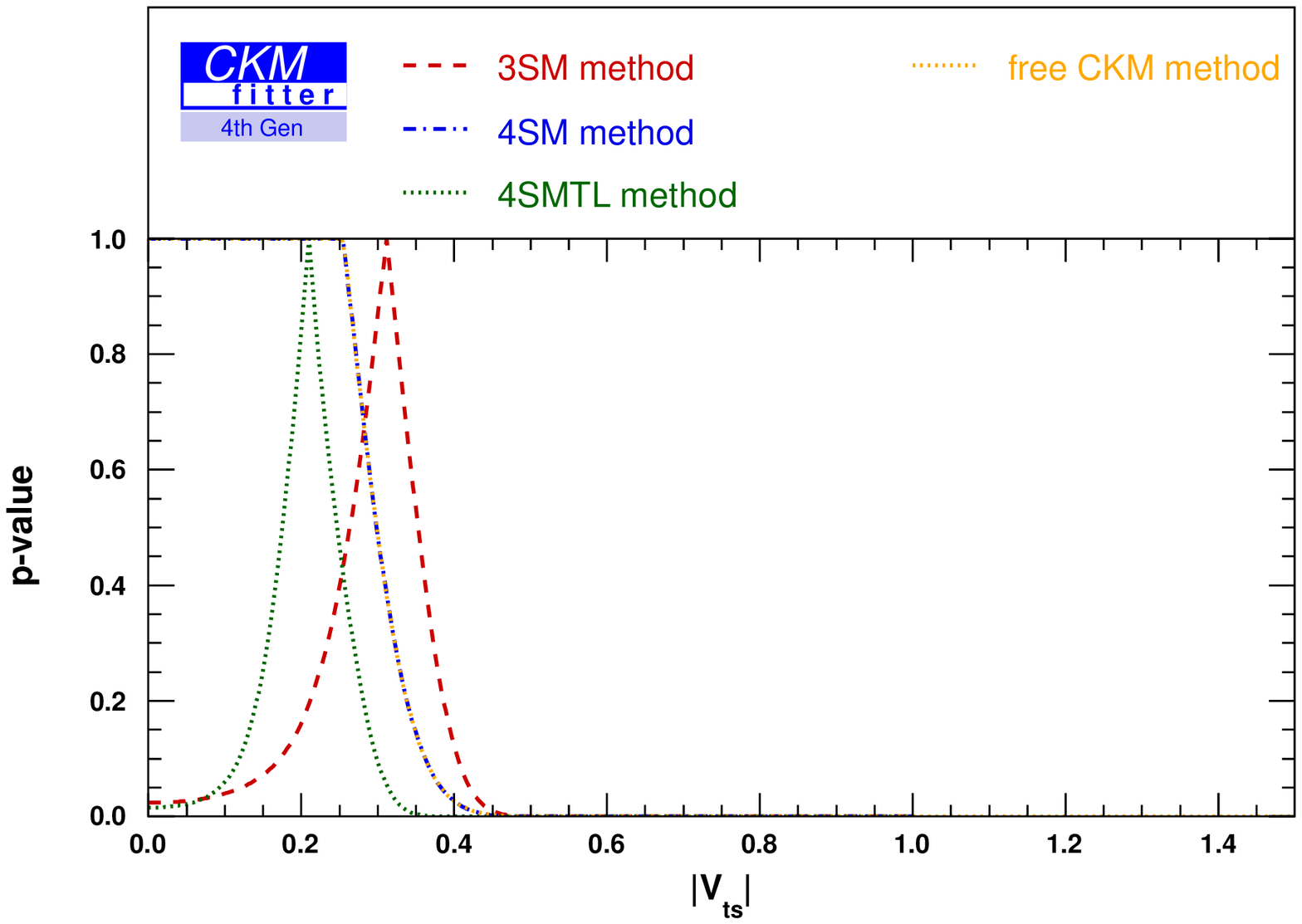}
}
\caption{Constraint on $|V_{ts}|$ from the single top analysis within 
         different scenarios using $N_{1 b jet}^{2 jets}=84.3 \pm 26.8$.
         For the different colour-coding we refer to the caption of Fig.~\ref{Vtb_normal}.}
\label{Vts_normal}       
\end{figure}

\begin{figure}
\resizebox{0.99\textwidth}{!}{
  \includegraphics{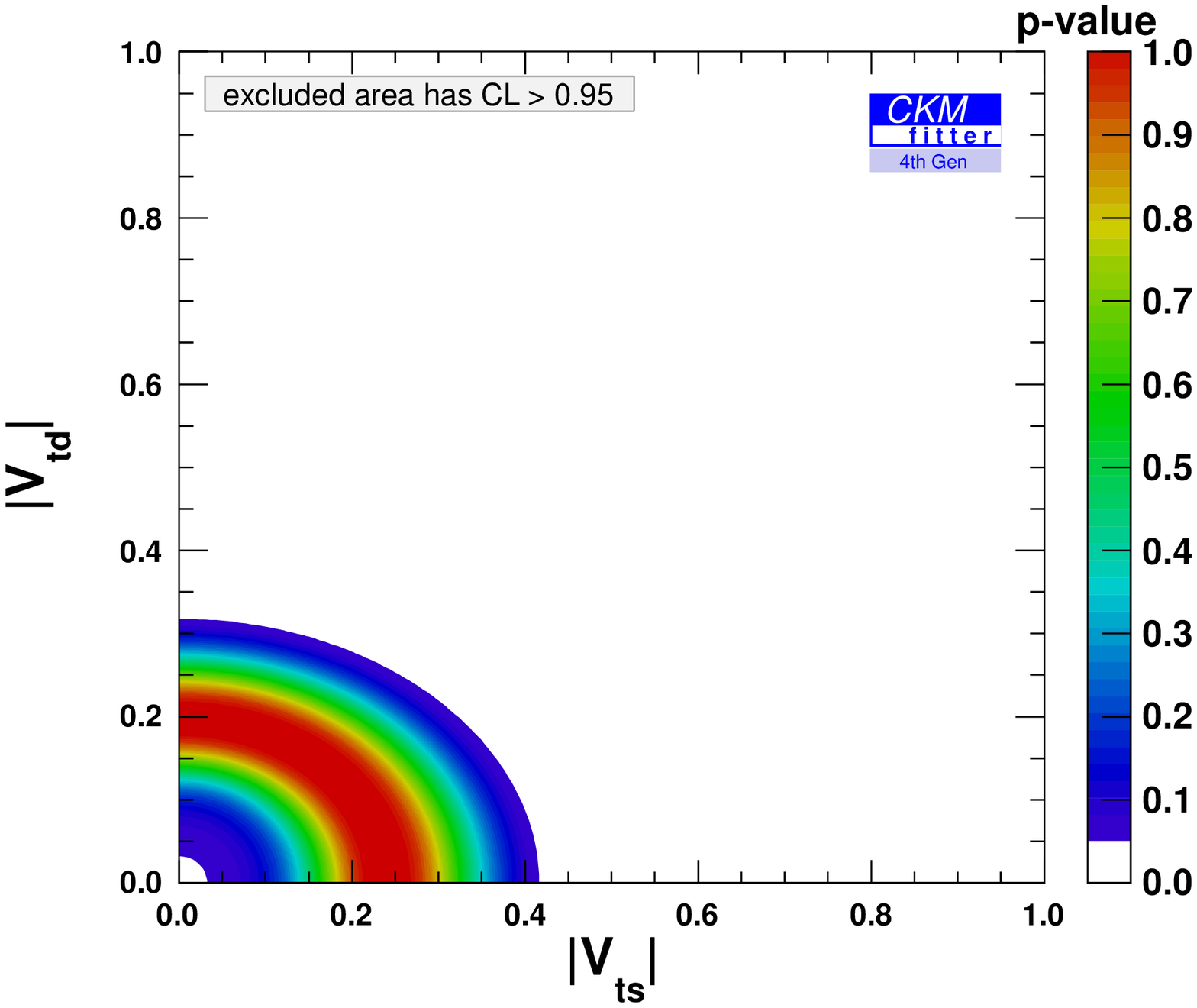}
}
\caption{The two-dimensional constraint on $|V_{td}|$ as a function of $|V_{ts}|$ 
         within the `4SM method'
         from the single top analysis using $N_{1 b jet}^{2 jets}=84.3 \pm 26.8$ 
         and $R=0.90 \pm 0.04$.}
\label{VtdVts_normal}       
\end{figure}

\begin{figure}
\resizebox{0.99\textwidth}{!}{
  \includegraphics{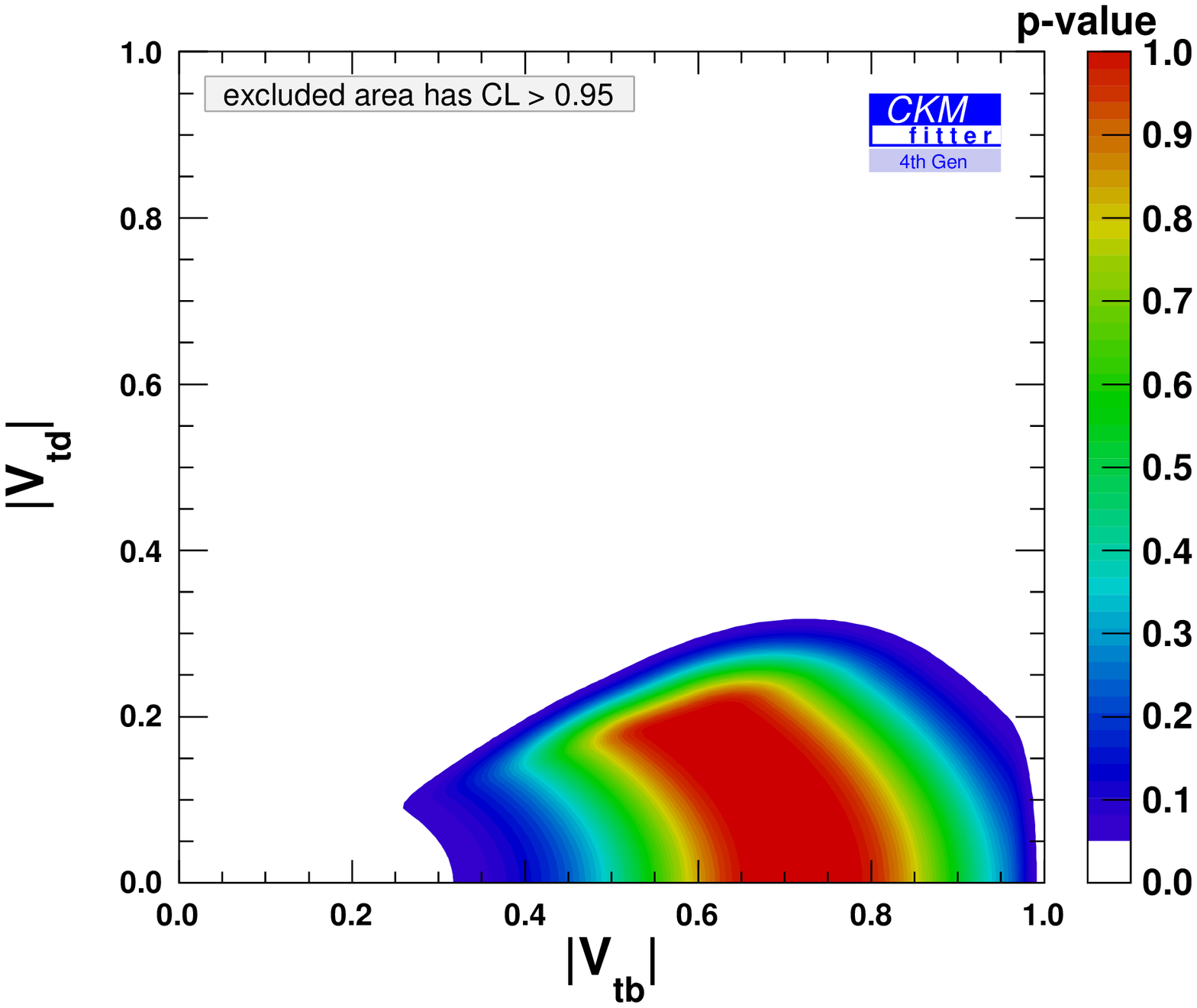}
}
\caption{The two-dimensional constraint on $|V_{td}|$ as a function of $|V_{tb}|$ 
         within the `4SM method'
         from the single top analysis using $N_{1 b jet}^{2 jets}=84.3 \pm 26.8$ 
         and $R=0.90 \pm 0.04$.}
\label{VtdVtb_normal}       
\end{figure}

\begin{figure}
\resizebox{0.99\textwidth}{!}{
  \includegraphics{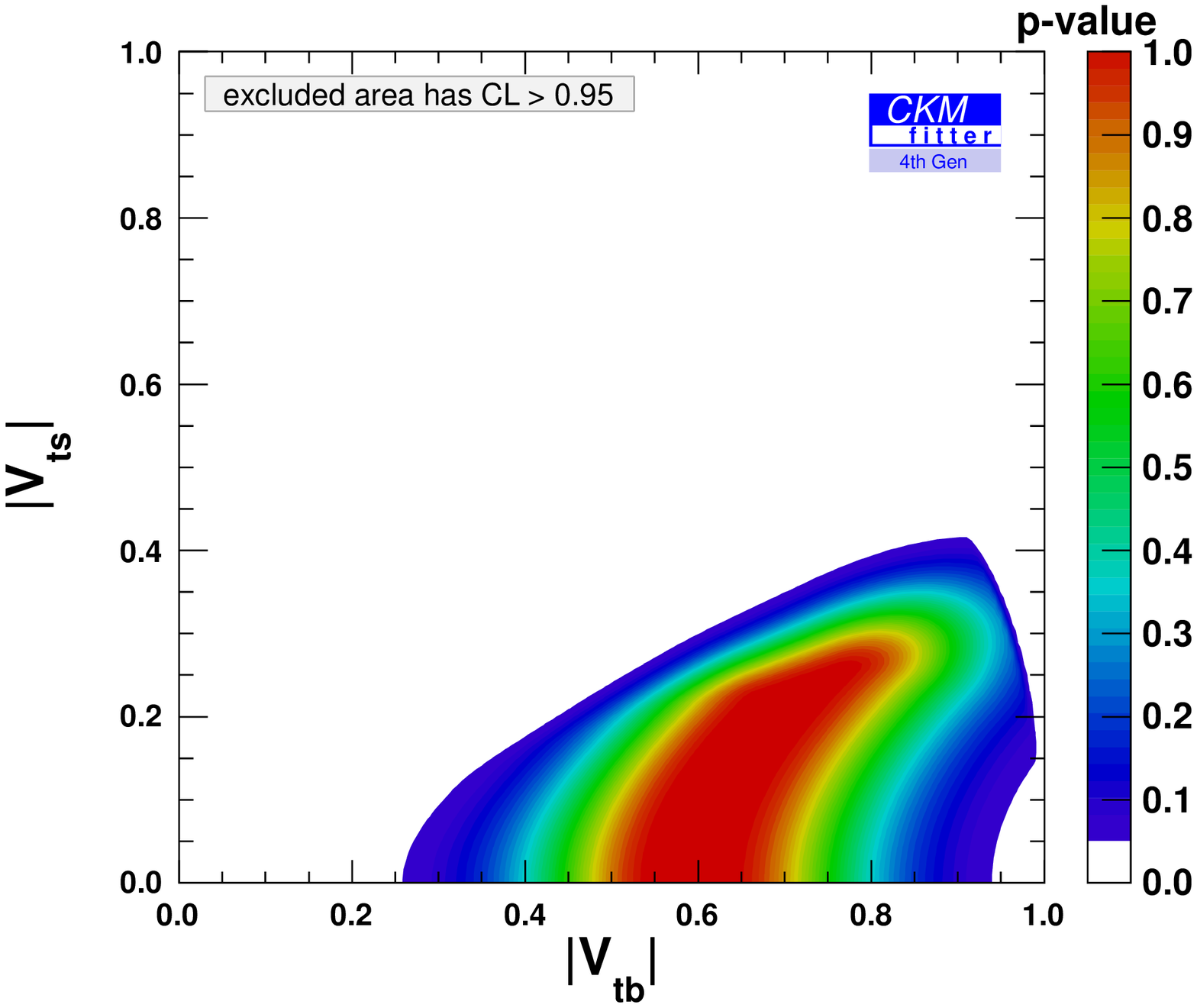}
}
\caption{The two-dimensional constraint on $|V_{ts}|$ as a function of $|V_{tb}|$ 
         within the `4SM method'
         from the single top analysis  using $N_{1 b jet}^{2 jets}=84.3 \pm 26.8$ 
         and $R=0.90 \pm 0.04$.}
\label{VtsVtb_normal}       
\end{figure}

\begin{figure}
\resizebox{0.99\textwidth}{!}{
  \includegraphics{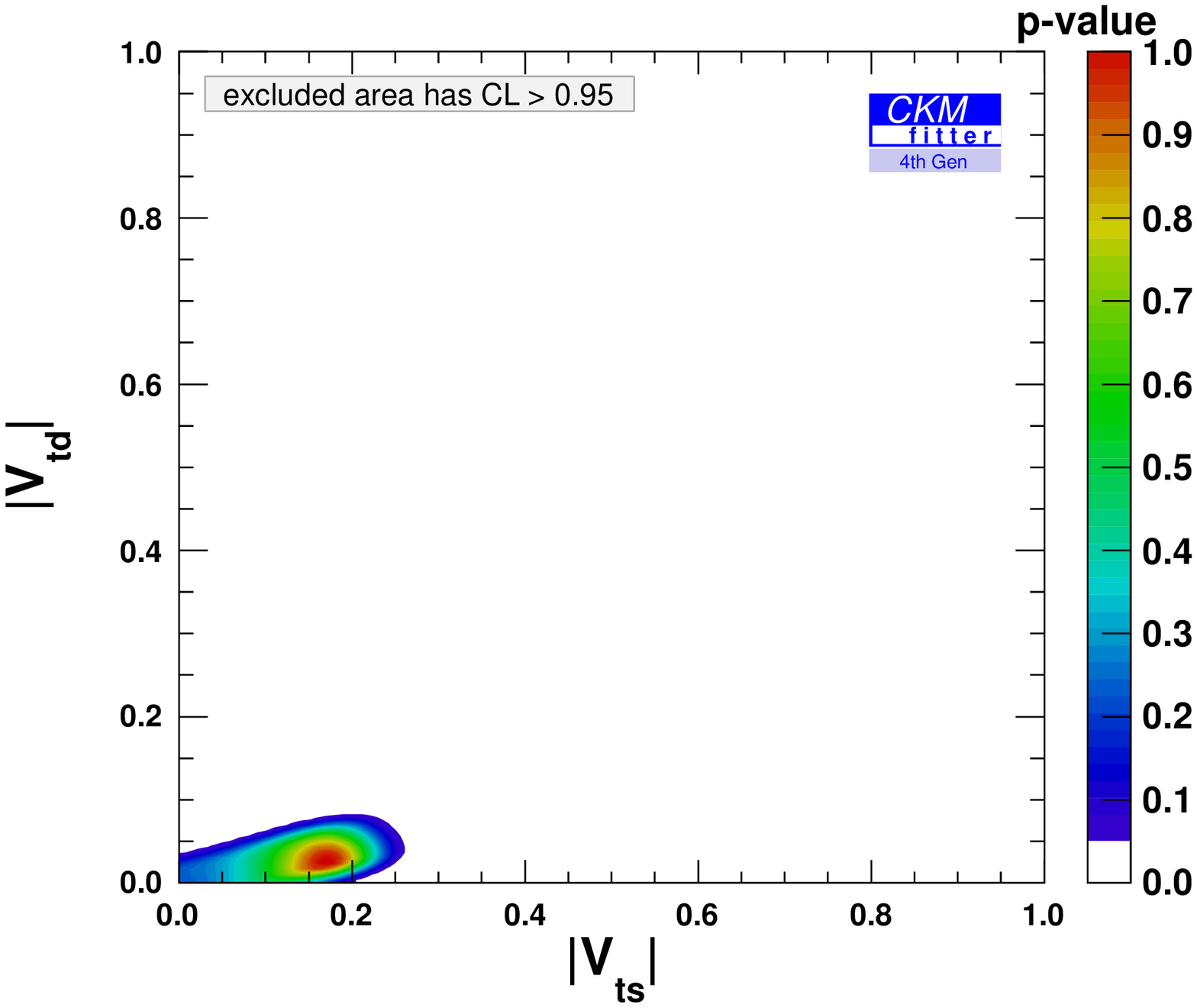}
}
\caption{The two-dimensional constraint on $|V_{td}|$ and $|V_{ts}|$ within 
         the `4SMTL method' using $N_{1 b jet}^{2 jets}=84.3 \pm 26.8$ and 
         $R=0.90 \pm 0.04$ together with constraints on $|V_{ud}|$, $|V_{us}|$, 
         $|V_{ub}|$, $|V_{cd}|$, 
         $|V_{cb}|$, and 
         ${\cal{B}}(W \rightarrow \ell \nu_{\ell})$ as explained in the text.}
\label{VtdVts_TL_normal}       
\end{figure}

\begin{figure}
\resizebox{0.99\textwidth}{!}{
  \includegraphics{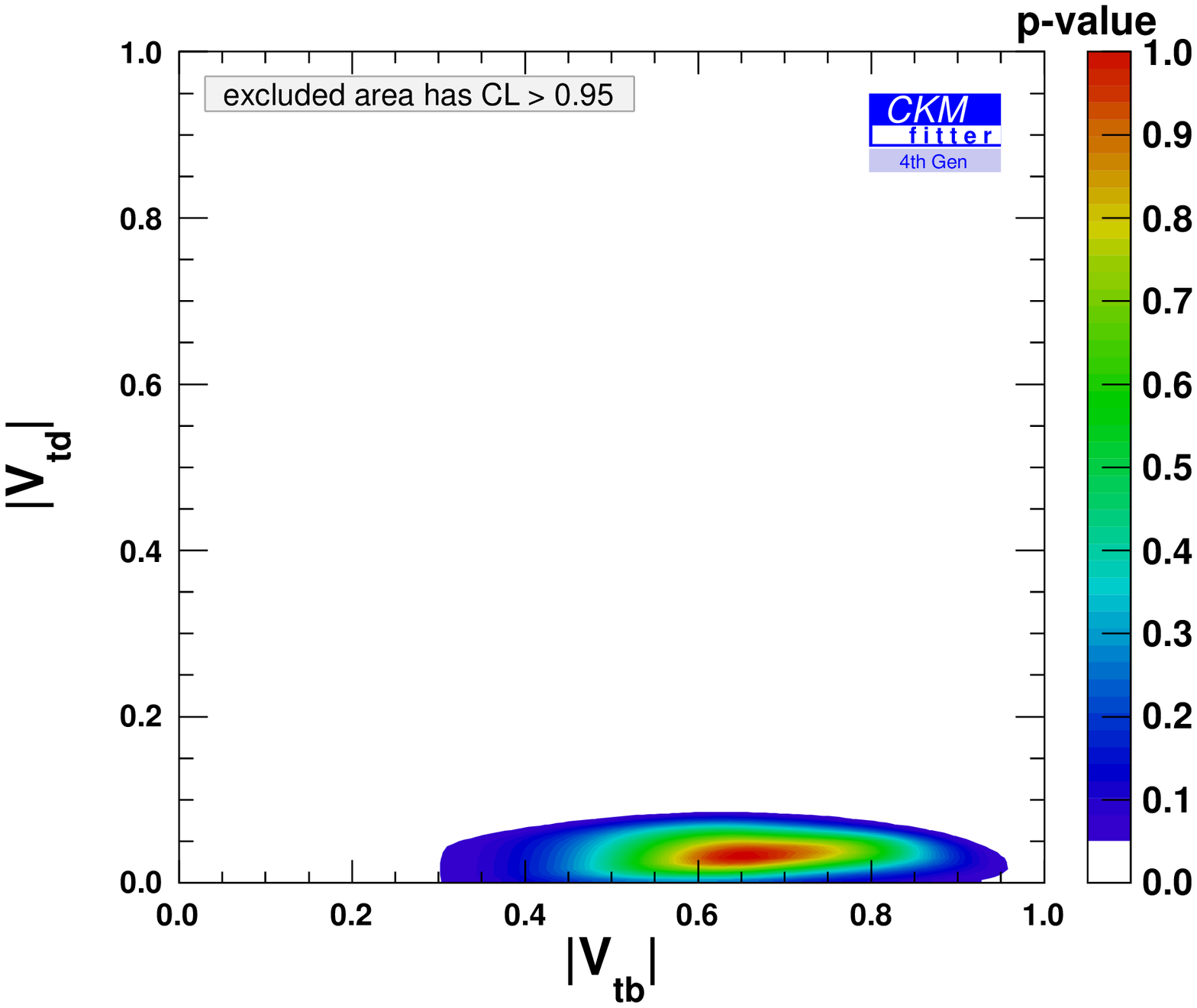}
}
\caption{The two-dimensional constraint on $|V_{td}|$ and $|V_{tb}|$ within 
         the `4SMTL method' using $N_{1 b jet}^{2 jets}=84.3 \pm 26.8$ and 
         $R=0.90 \pm 0.04$ together with constraints on $|V_{ud}|$, $|V_{us}|$, 
         $|V_{ub}|$, $|V_{cd}|$, 
         $|V_{cb}|$, and 
         ${\cal{B}}(W \rightarrow \ell \nu_{\ell})$ as explained in the text.}
\label{VtdVtb_TL_normal}       
\end{figure}

\begin{figure}
\resizebox{0.99\textwidth}{!}{
  \includegraphics{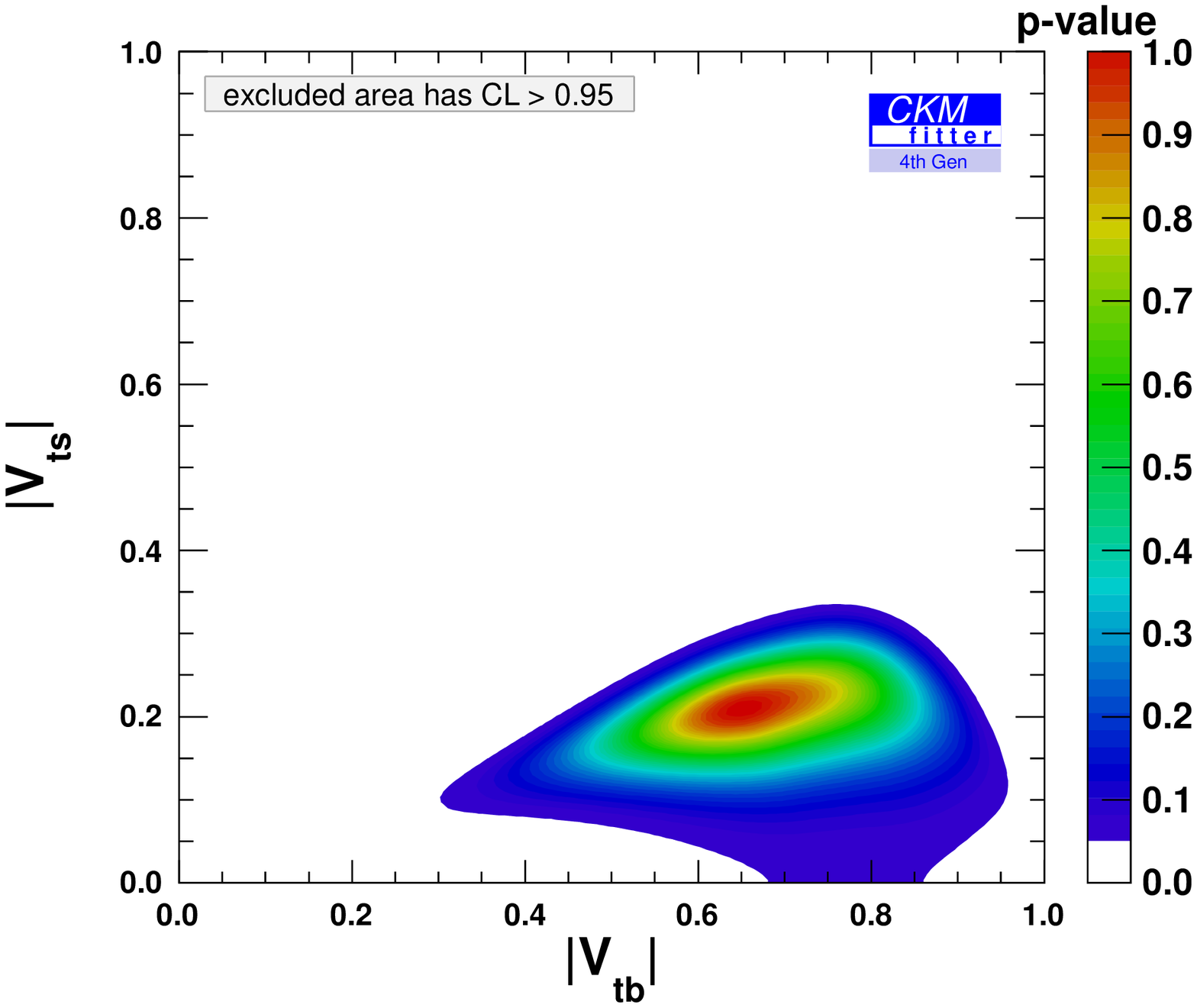}
}
\caption{The two-dimensional constraint on $|V_{ts}|$ and $|V_{tb}|$ within 
         the `4SMTL method' using $N_{1 b jet}^{2 jets}=84.3 \pm 26.8$ and 
         $R=0.90 \pm 0.04$ together with constraints on $|V_{ud}|$, $|V_{us}|$, 
         $|V_{ub}|$, $|V_{cd}|$, 
         $|V_{cb}|$, and 
         ${\cal{B}}(W \rightarrow \ell \nu_{\ell})$ as explained in the text.}
\label{VtsVtb_TL_normal}       
\end{figure}

\begin{figure}
\resizebox{0.99\textwidth}{!}{
  \includegraphics{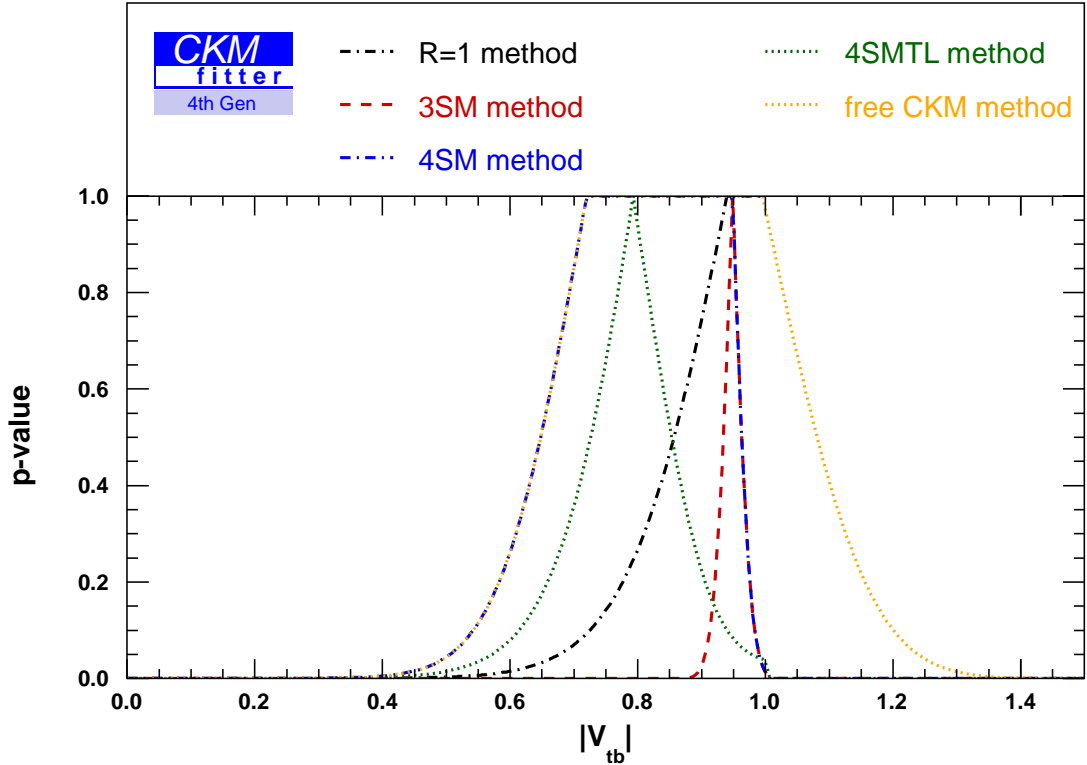}
}
\caption{Constraint on $|V_{tb}|$ from a single top analysis assuming a 
         hypothetical measurement of $N_{1bjet}^{2jets}=142.8 \pm 34.6$:
         black dashed-dotted curve: p-value obtained in the `$R=1$ method';
         red dashed curve: p-value obtained in the `3SM method' setting $R=0.90 \pm 0.04$;
         blue dashed-dotted curve: p-value obtained in the `4SM method' setting $R=0.90 \pm 0.04$;
         green dotted curve: p-value obtained in the `4SMTL method' setting $R=0.90 \pm 0.04$
         and using in addition constraints on $|V_{ud}|$, $|V_{us}|$, $|V_{ub}|$, 
         $|V_{cd}|$, 
         $|V_{cb}|$, and ${\cal{B}}(W \rightarrow \ell \nu_{\ell})$;
         orange dotted curve: p-calue obtained in the `free CKM method' setting $R=0.90 \pm 0.04$.}
\label{VtbKombi_R090_Nbjet141}       
\end{figure}

Once the tree-level measurements of $|V_{ud}|$, $|V_{us}|$, $|V_{ub}|$, $|V_{cd}|$, 
$|V_{cb}|$, and the measurements of ${\cal{B}}(W \rightarrow \ell \nu_{\ell})$ 
are added (`4SMTL method'), the constraint on $|V_{tb}|$ (green dotted line) 
becomes much tighter and even tighter than the one from the `$R=1$ method'. 
Due to the deviation of the measured $R$ value from one the 4SMTL constraint 
is significantly shifted with respect to the standard $|V_{tb}|$ value.
We also present the constraints on $|V_{td}|$ and $|V_{ts}|$ in 
Figs.~\ref{Vtd_normal} and~\ref{Vts_normal}.
Figs.~\ref{Vtd_normal} and~\ref{Vts_normal} show the corresponding constraints 
on $|V_{td}|$ and $|V_{ts}|$. In this case, the `$R=1$ method' does not appear
as, by definition, both CKM elements are forced to be zero.\\
To understand better the constraints on $|V_{td}|$ and $|V_{ts}|$ it is 
instructive to study the correlations between the constraints on $|V_{td}|$, 
$|V_{ts}|$, and $|V_{tb}|$. In Figs.~\ref{VtdVts_normal},~\ref{VtdVtb_normal},
and~\ref{VtsVtb_normal}, we show two-dimensional constraints for the `4SM method'.
The ring-like constraint in the $|V_{td}|$-$|V_{ts}|$ plane is driven by the
$R$ measurement. One does not obtain though a perfect circular shape as the 
single top yield has different sensitivities to $|V_{td}|$ and $|V_{ts}|$ due to 
the different sizes of the $t$-channel cross sections and efficiencies.
For the same reason, the $|V_{td}|$-$|V_{tb}|$ plane shows an anti-correlation
and the $|V_{ts}|$-$|V_{tb}|$ a correlation.
In Figs.~\ref{VtdVts_TL_normal},~\ref{VtdVtb_TL_normal}, and~\ref{VtsVtb_TL_normal} 
we present the corresponding constraints in the `4SMTL method'. Due to the very 
precisely measured value of $|V_{ud}|$, the allowed values of $|V_{td}|$ are 
well-constrained to be below $0.08$. As a consequence, the devation of $R$ from 1
can only be compensated by a rather large value of $|V_{ts}|$ of order 0.21 which
is still allowed by the unitarity constraints in the second column of the 
$4 \times 4$ CKM matrix but not preferred. 
Next, we study the constraints in a hypothetical scenario where the measured 
single top yield is in perfect agreement with the expected number of signal 
events in the 3SM scenario assuming $|V_{tb}|=1$ but still using the measured 
$R$ value. In this hypothetical scenario we set $N_{1bjet}^{2jets}= 142.8 \pm 34.6$. 
The corresponding plot for the constraint on $|V_{tb}|$ is shown in 
Fig.~\ref{VtbKombi_R090_Nbjet141}. In this case, although the measured single 
top yield perfectly fits with the 3SM expectation the $|V_{tb}|$ constraints 
in the 3SM, 4SM and even `4SMTL method's still deviate significantly from the 
standard scenario.

\subsection{Numerical results with a modified $R$ input}

\begin{figure}
\resizebox{0.99\textwidth}{!}{
  \includegraphics{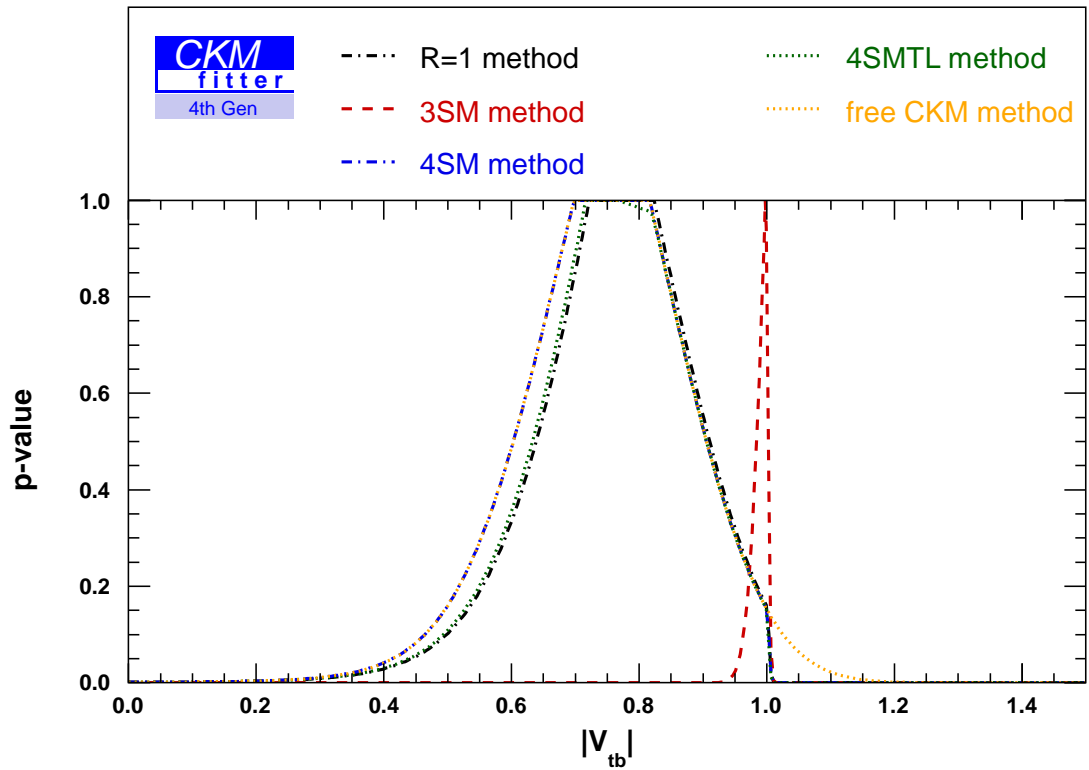}
}
\caption{Constraint on $|V_{tb}|$ from a single top analysis using $N_{1bjet}^{2jets}=84.3 \pm 26.8$:
         black dashed-dotted curve: p-value obtained in the `$R=1$ method';
         red dashed curve: p-value obtained in the `3SM method' setting $R=0.99 \pm 0.04$;
         blue dashed-dotted curve: p-value obtained in the `4SM method' setting $R=0.99 \pm 0.04$;
         green dotted curve: p-value obtained in the `4SMTL method' setting $R=0.99 \pm 0.04$
         and using in addition constraints on $|V_{ud}|$, $|V_{us}|$, $|V_{ub}|$, 
         $|V_{cd}|$, 
         $|V_{cb}|$, and ${\cal{B}}(W \rightarrow \ell \nu_{\ell})$;
         orange dotted curve: p-calue obtained in the `free CKM method' setting $R=0.99 \pm 0.04$.}
\label{VtbKombi_R099}       
\end{figure}

\begin{figure}
\resizebox{0.99\textwidth}{!}{
  \includegraphics{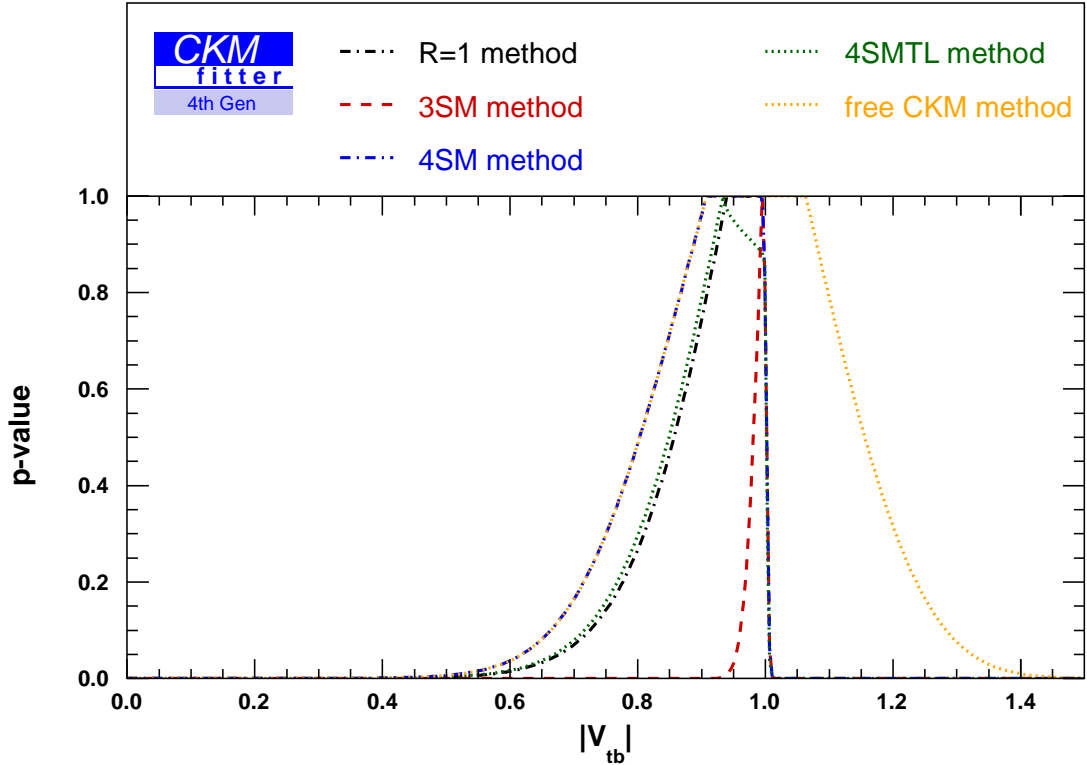}
}
\caption{Constraint on $|V_{tb}|$ from a single top analysis assuming
         hypothetical measurement $N_{1bjet}^{2jets}=142.8 \pm 34.6$:
         black dashed-dotted curve: p-value obtained in the `$R=1$ method';
         red dashed curve: p-value obtained in the `3SM method' setting $R=0.99 \pm 0.04$;
         blue dashed-dotted curve: p-value obtained in the `4SM method' setting $R=0.99 \pm 0.04$;
         green dotted curve: p-value obtained in the `4SMTL method' setting $R=0.99 \pm 0.04$
         and using in addition constraints on $|V_{ud}|$, $|V_{us}|$, $|V_{ub}|$, 
         $|V_{cd}|$, 
         $|V_{cb}|$, and ${\cal{B}}(W \rightarrow \ell \nu_{\ell})$;
         orange dotted curve: p-calue obtained in the `free CKM method' setting $R=0.99 \pm 0.04$.}
\label{VtbKombi_R099_Nbjet141}       
\end{figure}

We now perform an instructive exercise. As the new $R$ measurement from 
D0 leads to a large $\chi^{2}$ value in the 4SM fit when combined with 
tree-level constraints and the single top measurements we modify the 
central value of $R$ such that one obtains a reasonably small $\chi^{2}$, 
that is, the $R$ measurement would be in agreement with the tree-level 
constraints and the measured single top yield.
We choose $R = 0.99 \pm 0.04$ and present the $|V_{tb}|$ constraint
in Figs.~\ref{VtbKombi_R099} and \ref{VtbKombi_R099_Nbjet141} corresponding
to measured event yields of $N_{1bjet}^{2jets}=84.3 \pm 26.8$, respectively, 
$N_{1bjet}^{2jets}=142.8 \pm 34.6$.
Once $R$ is very close to one the difference in the constraints on $|V_{tb}|$
between the `$R=1$ method and the `4SM(TL) method' becomes small (tiny).

\subsection{The top-quark decay width}
Under the assumption that the top-quark decays have to proceed through 
$t \rightarrow q + W$, $q = d, s, b$, the total top-quark decay width for 
massless final-state quarks at NLO in QCD is given by~\cite{TopWidth}
\begin{eqnarray}\label{Eq:TopWidth}
\Gamma_{t} &=& (|V_{td}|^2 + |V_{ts}|^2 +|V_{tb}|^2) \frac{G_{F} m_{t}^{3}}{8 \pi \sqrt{2}} 
             (1 - \frac{m_{W}^{2}}{m_{t}^{2}})^{2} (1 + 2 \frac{m_{W}^{2}}{m_{t}^{2}}) \nonumber \\
           &\times& \left[1 - \frac{2 \alpha_{s}}{3 \pi} (\frac{2 \pi^{2}}{3} - \frac{5}{2} ) \right],
\end{eqnarray}
where $G_{F}=1.16637 \cdot 10^{-5} \,{\rm GeV}^{-2}$~\cite{PDG2010} is the Fermi 
constant, and  the QCD corrections are included in the $m_{W}/m_{t} \ll 1$ limit.
\begin{figure}
\resizebox{0.99\textwidth}{!}{
  \includegraphics{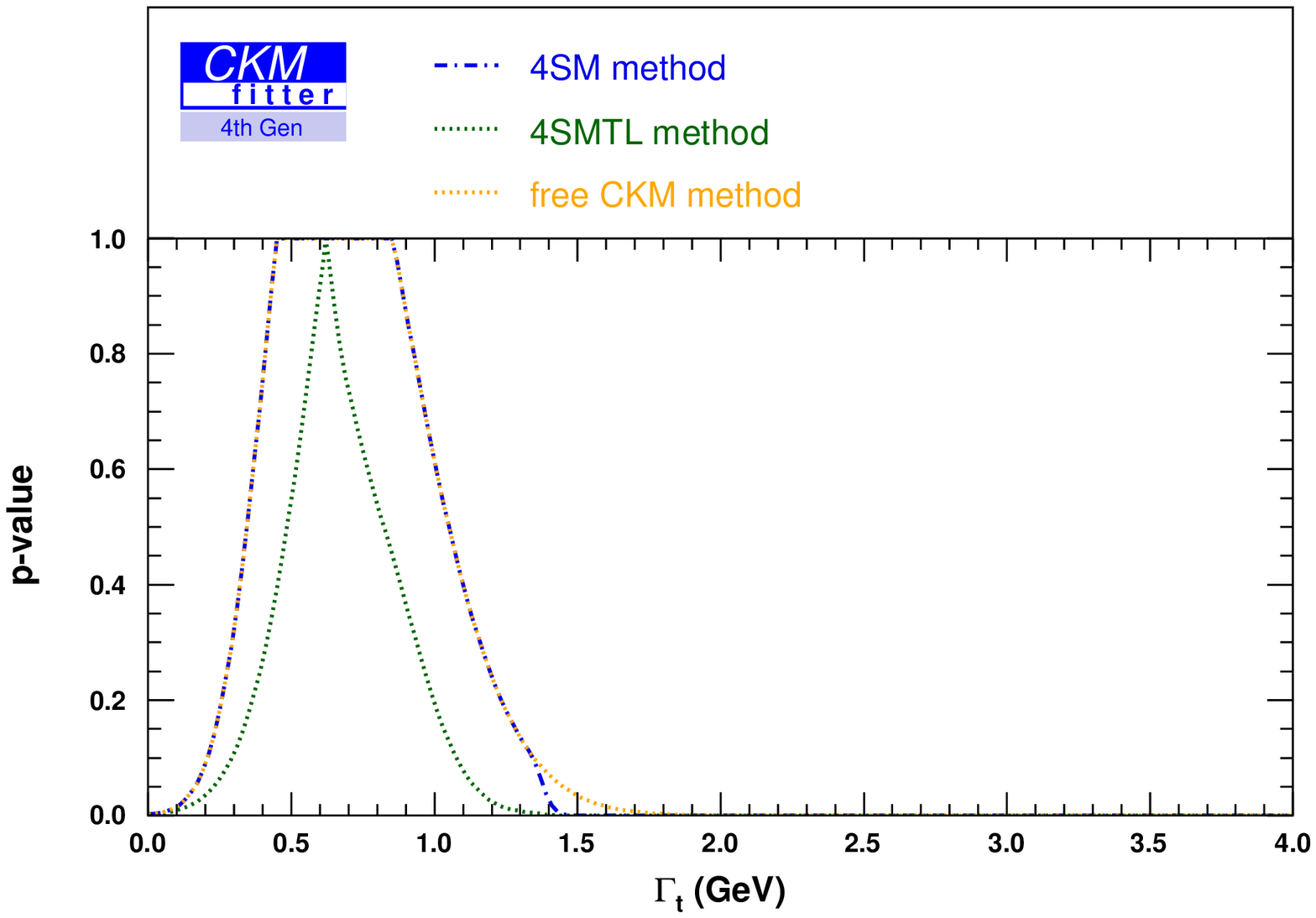}
}
\caption{Constraint on $\Gamma_{t}$ from a single top analysis assuming
         hypothetical measurement $N_{1bjet}^{2jets}=84.3 \pm 26.8$:
         blue dashed-dotted curve: p-value obtained in the `4SM method' setting $R=0.90 \pm 0.04$;
         green dotted curve: p-value obtained in the `4SMTL method' setting $R=0.90 \pm 0.04$
         and using in addition constraints on $|V_{ud}|$, $|V_{us}|$, $|V_{ub}|$, 
         $|V_{cd}|$, 
         $|V_{cb}|$, and ${\cal{B}}(W \rightarrow \ell \nu_{\ell})$;
         orange dotted curve: p-calue obtained in the `free CKM method' setting $R=0.90 \pm 0.04$.}
\label{TopWidthKombi_R090}       
\end{figure}
\begin{figure}
\resizebox{0.99\textwidth}{!}{
  \includegraphics{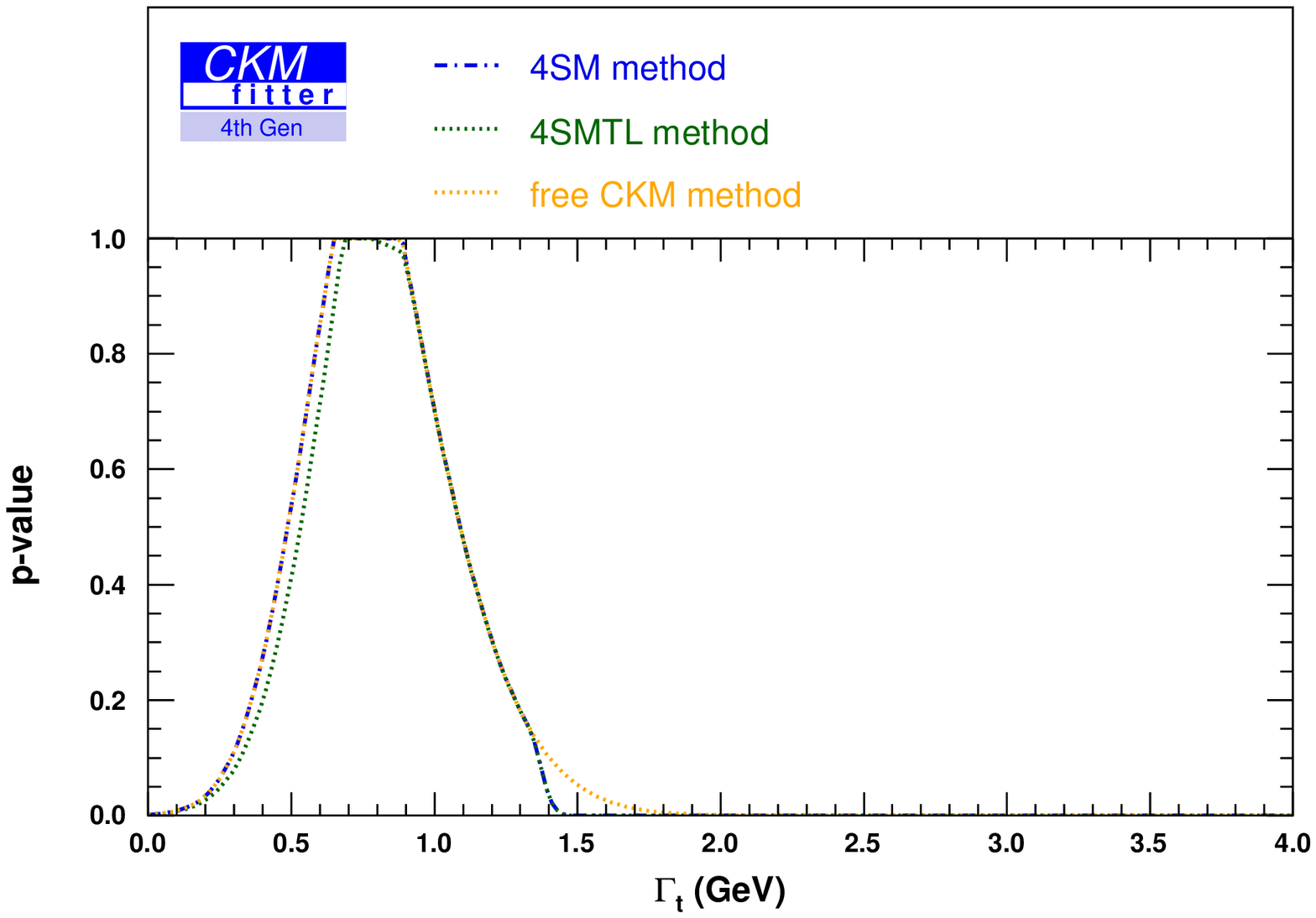}
}
\caption{Constraint on $\Gamma_{t}$ from a single top analysis assuming
         hypothetical measurement $N_{1bjet}^{2jets}=84.3 \pm 26.8$:
         blue dashed-dotted curve: p-value obtained in the `4SM method' setting $R=0.99 \pm 0.04$;
         green dotted curve: p-value obtained in the `4SMTL method' setting $R=0.99 \pm 0.04$
         and using in addition constraints on $|V_{ud}|$, $|V_{us}|$, $|V_{ub}|$, 
         $|V_{cd}|$, 
         $|V_{cb}|$, and ${\cal{B}}(W \rightarrow \ell \nu_{\ell})$;
         orange dotted curve: p-calue obtained in the `free CKM method' setting $R=0.99 \pm 0.04$.}
\label{TopWidthKombi_R099}       
\end{figure}
\begin{figure}
\resizebox{0.99\textwidth}{!}{
  \includegraphics{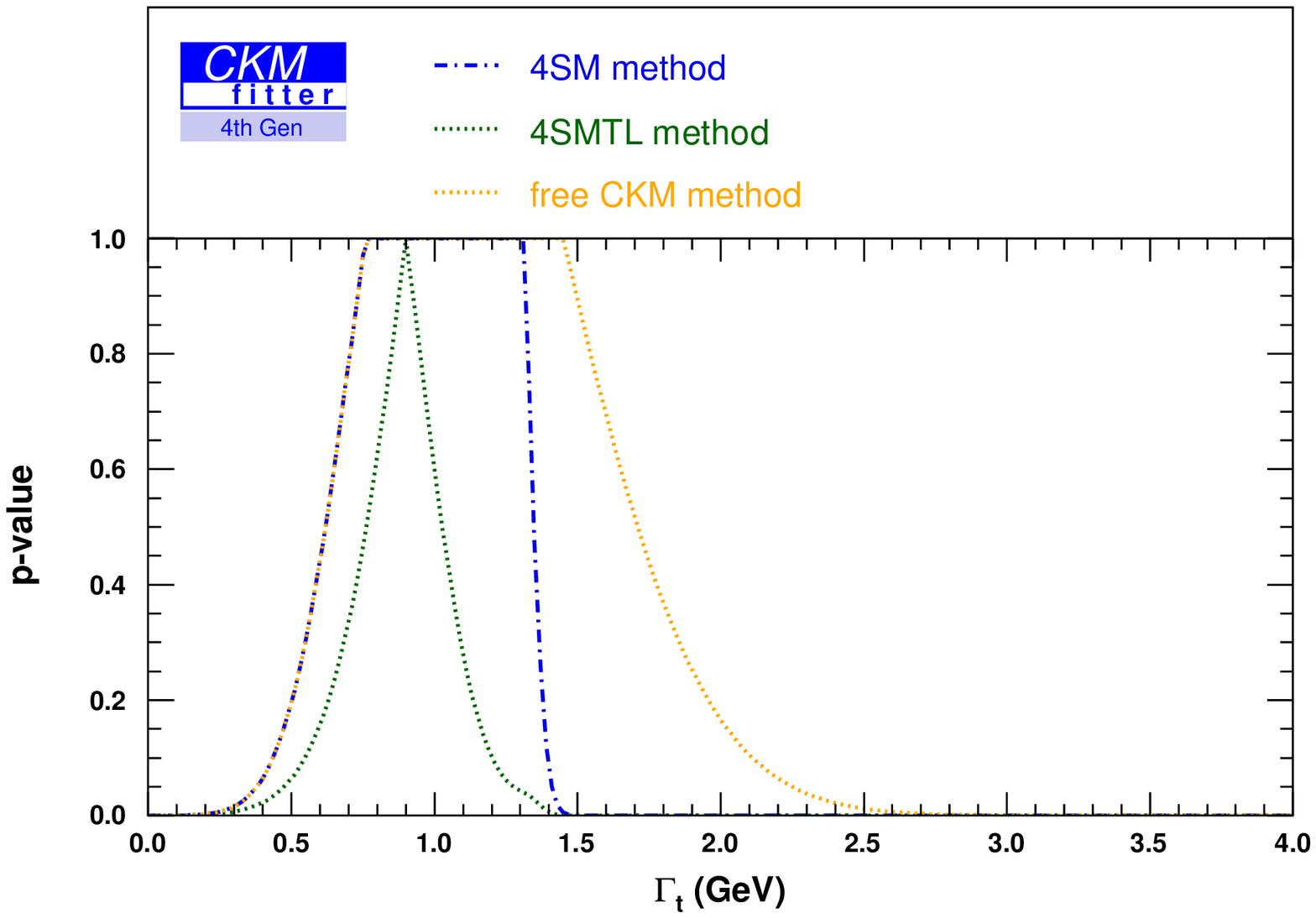}
}
\caption{Constraint on $\Gamma_{t}$ from a single top analysis assuming
         hypothetical measurement $N_{1bjet}^{2jets}=142.8 \pm 34.6$:
         blue dashed-dotted curve: p-value obtained in the `4SM method' setting $R=0.90 \pm 0.04$,
         green dotted curve: p-value obtained in the `4SMTL method' setting $R=0.90 \pm 0.04$
         and using in addition constraints on $|V_{ud}|$, $|V_{us}|$, $|V_{ub}|$, 
         $|V_{cd}|$, 
         $|V_{cb}|$, and ${\cal{B}}(W \rightarrow \ell \nu_{\ell})$;
         orange dotted curve: p-calue obtained in the `free CKM method' setting $R=0.90 \pm 0.04$.}
\label{TopWidthKombi_R090_N1bjet141}       
\end{figure}
\begin{figure}
\resizebox{0.99\textwidth}{!}{
  \includegraphics{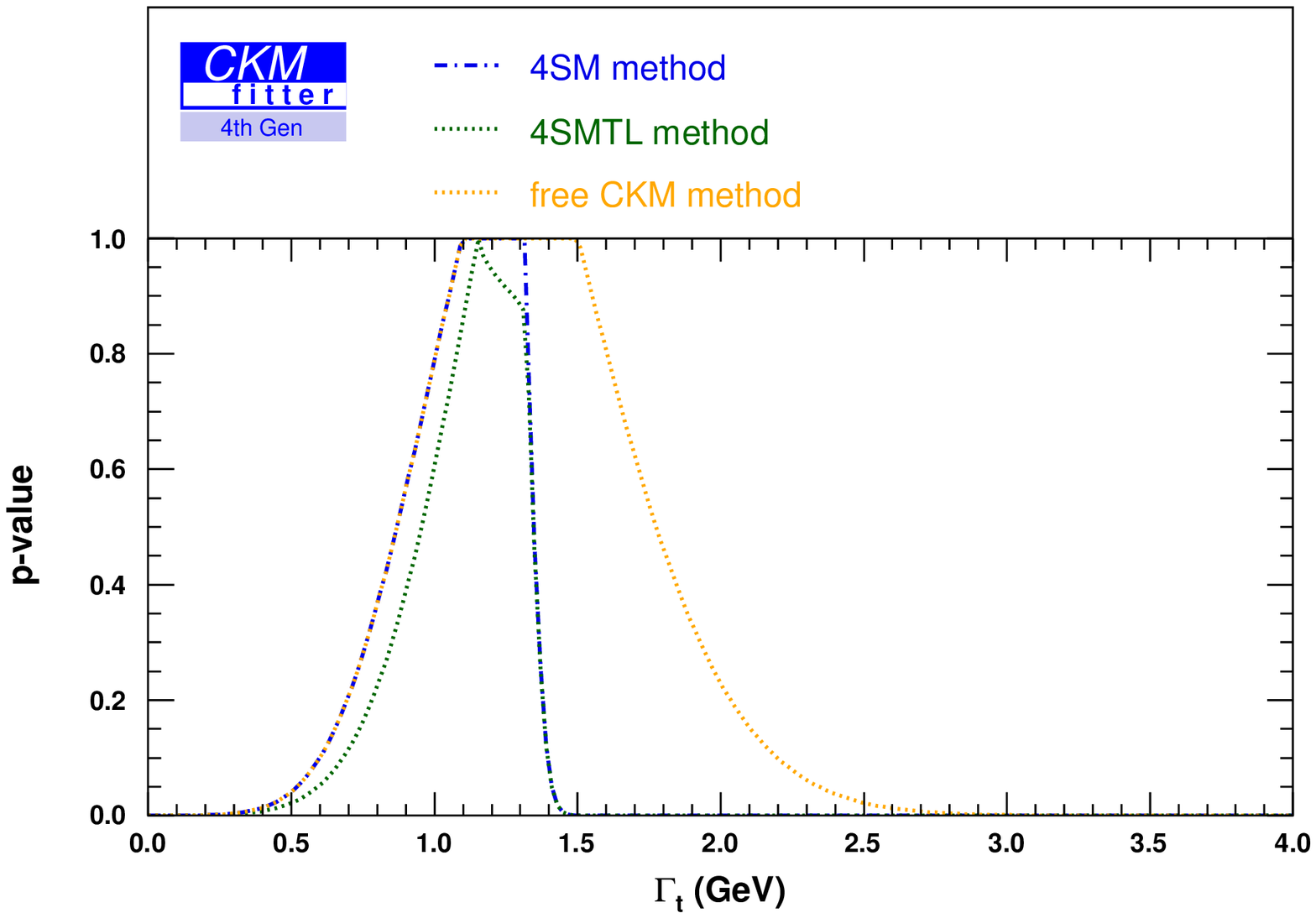}
}
\caption{Constraint on $\Gamma_{t}$ from a single top analysis assuming
         hypothetical measurement $N_{1bjet}^{2jets}=142.8 \pm 34.6$:
         blue dashed-dotted curve: p-value obtained in the `4SM method' setting $R=0.99 \pm 0.04$;
         green dotted curve: p-value obtained in the `4SMTL method' setting $R=0.99 \pm 0.04$
         and using in addition constraints on $|V_{ud}|$, $|V_{us}|$, $|V_{ub}|$, 
         $|V_{cd}|$, 
         $|V_{cb}|$, and ${\cal{B}}(W \rightarrow \ell \nu_{\ell})$;
         orange dotted curve: p-calue obtained in the `free CKM method' setting $R=0.99 \pm 0.04$.}
\label{TopWidthKombi_R099_N1bjet141}       
\end{figure}
In the 3SM, $|V_{td}|^2 + |V_{ts}|^2 + |V_{tb}|^2$ is equal to one due to 
unitarity and one predicts the top-quark width to be $\Gamma_{t} = (1.32 \pm 0.04)\,{\rm GeV}$ 
where we have used $\alpha_{s}=0.118$, $m_{W}= 80.399~{\rm GeV}$ for the $W$-boson mass, 
and $m_{t}=(172.0 \pm 1.6)\,{\rm GeV}$ for the top-quark mass as in our numerical analysis.\\
In Figs.~\ref{TopWidthKombi_R090},~\ref{TopWidthKombi_R099},~\ref{TopWidthKombi_R090_N1bjet141},
and~\ref{TopWidthKombi_R099_N1bjet141}, 
we present the constraints on $\Gamma_{t}$ obtained in the `4SM method', `4SMTL method', 
and the `free CKM method' respectively. The latter method can be considered the closest 
to that employed by the D0 collaboration to extract $\Gamma_{t}$. The constraints are 
determined for the different values of $R$ and $N_{1bjet}^{2jets}$ used in the numerical analyses 
described above.
As expected, the constraints on $\Gamma_{t}$ are identical for the `4SM method'
and the `free CKM method' for values of $\Gamma_{t}$ below the 3SM expectation
of $\Gamma_{t} = 1.32{\rm GeV}$ as this is the allowed region of $\Gamma_{t}$
for $|V_{td}|^2 + |V_{ts}|^2 + |V_{tb}|^2 \le 1$.\\
Analoguous to the $|V_{tq}|$ constraints the differences between the `4SM method'
and the `4SMTL method' turn out to be significant for $R=0.90 \pm 0.04$.

\section{Conclusion}
We have presented a strategy to extract the CKM matrix elements $|V_{tb}|$ and
also $|V_{td}|$ and $|V_{ts}|$
from single top production measurements at the Tevatron which goes beyond 
the `$R=1$ method' assuming $|V_{tb}| \gg |V_{td}|,|V_{ts}|$, and 
explicitely takes into account $d$- and $s$-quark contributions to the $s$- 
and $t$-channel production and to the decay of top quarks. The method provides 
information that can be directly used to put constraints on 4SM and other 
scenarios with new heavy quarks and to extract the top-quark width within 
these scenarios.

For sake of illustration, we applied the method to CDF data on lepton $+$ 
missing transverse energy $+$ two jet events with one reconstructed $b$-jet 
and to recent D0 results on the top branching ratio to $b$-quarks. 
We estimated the relevant efficiencies from a leading order Monte-Carlo 
simulation.
The constraints within a 4SM scenario from the single top measurements and 
from $R$ can only be combined with other flavor observables in a consistent 
way within a global analysis. As an example, we studied a global analysis of 
the single top yield $N_{1 b jet}^{2 jets}$ and $R$ with tree-level flavor 
measurements to constrain CKM matrix elements in a 4SM scenario. 

Our simplified analysis shows that with the recent measurement of $R=0.90 \pm 0.04$ 
presented by the D0 collaboration the constraint on $|V_{tb}|$ in a 4SM scenario 
differs significantly from the `$R=1$ method' even if $|V_{td}|$ and 
$|V_{ts}|$ are constrained by tree-level measurements of $|V_{ud}|$, $|V_{us}|$, 
$|V_{ub}|$, $|V_{cd}|$, 
$|V_{cb}|$, and by leptonic $W$-decays thanks to $4 \times 4$ unitarity. 

Further valuable and accurate information could be easily extracted from the 
Tevatron data by including also event rates with two or three jets, both of 
them $b$-tagged and by employing NLO MC + data validation for the determination 
of the efficiencies. In particular, the fact that $s$-channel and $t$-channel
at the Tevatron are expected to give comparable rates of events in the SM 
provides a leverage that has no equivalent at the LHC and put CDF and D0 in 
a very competitive position. 

While this paper concentrates on single top-quark production we would like 
to point out that a value of $R$ being significantly smaller than one might 
have important implications for $t \bar{t}$ production measurements. 
Without taking into account $R$ being smaller than one, the measured $t \bar{t}$ 
cross section would underestimate the true cross-section value which in turn 
would overestimate the top-quark mass extracted from the cross section measurement. 
As an example, if a $t \bar{t}$ cross-section measurement performed at Tevatron 
used one b-tagged jet, then the cross section would be underestimated by a factor 
$R$. 
Correspondingly, the extracted top-quark mass would be overestimated by about 
$O(3~{\rm GeV})$ which can be read off e.g. from Ref.~\cite{Langenfeld:2010zz} 
where the top-quark mass extraction from $t \bar{t}$ cross-section measurements 
is discussed in detail.
This issue might become relevant when comparing the top-quark mass extracted 
from a cross-section measurement with the one from direct measurements.

The method outlined in this work can also be applied to single top-quark 
measurements at the LHC. In this case, however, $s$-channel single top 
production is very small, while $Wt$ associated production becomes visible 
and therefore could be included. Furthermore, at the LHC the $t$-channel 
rate of top and anti-top is different due to the $pp$ initial state, 
$d$'s are valence (+sea) quarks while $\bar d$ are only sea quarks.
Contributions of the $d$ from the $s$ contributions could therefore be 
singled out by an accurate charge asymmetry measurement. Rapidity distributions
could also provide a further handle~\cite{AguilarSaavedra:2010wf}.
At LHC, however, one expects a lower sensitivity to $d$- and $s$-contributions
since the $t$-channel cross sections for $d$- and $s$-contributions do not
differ from the $b$-contribution as much as at the Tevatron. As an example,
we quote for a center-of-mass energy of $8~{\rm TeV}$ the NLO cross-sections
of single top and single antitop production at the LHC in 
Tables~\ref{TABLE_XS_LHC_Top} and~\ref{TABLE_XS_LHC_Antitop} calculated as
the ones for Tevatron described in Sec.~\ref{Sec:Inputs_SingleTopandR}.
\begin{table}[Htp]
\renewcommand{\arraystretch}{1.3}
\centering
\begin{tabular}{|c|c|c|c|c|c|}\hline
Cross section    & value & PDF unc. & scale unc. \\
         &  [pb] &   [pb]   &   [pb] \\
\hline
$\sigma_{d}^{t}$ & $396.3$    & $\pm 15.1$    & $^{+2.8}_{-1.2}$ \\
$\sigma_{s}^{t}$ & $124.9$    & $\pm  7.4$    & $^{+1.5}_{-0.6}$ \\
$\sigma_{b}^{t}$ & $ 54.8$    & $\pm  0.7$    & $^{+1.8}_{-1.0}$ \\
\hline
\end{tabular}
\caption[TABLE_XS_LHC_Top]
{NLO cross section for $t$-channel single top production at the LHC for a 
 center-of-mass energy $\sqrt{s}=8~{\rm TeV}$ calculated with MCFM~\cite{MCFM} 
 using the PDF sets taken from MSTW2008~\cite{MSTW2008}.
}
\label{TABLE_XS_LHC_Top}
\end{table}
\begin{table}[Htp]
\renewcommand{\arraystretch}{1.3}
\centering
\begin{tabular}{|c|c|c|c|c|c|}\hline
Cross section    & value & PDF unc. & scale unc. \\
                 &  [pb] &  [pb]    &    [pb]    \\
\hline
$\sigma_{d}^{t}$ & $108.9$    & $\pm 5.0$     & $^{+1.5}_{-0.1}$ \\
$\sigma_{s}^{t}$ & $69.6$     & $\pm 3.5$     & $^{+1.2}_{-0.3}$ \\
$\sigma_{b}^{t}$ & $30.0$     & $\pm 0.6$     & $^{+0.9}_{-0.7}$ \\
\hline
\end{tabular}
\caption[TABLE_XS_LHC_Antitop]
{NLO cross section for $t$-channel single anti-top production at the 
 LHC for a center-of-mass energy $\sqrt{s}=8~{\rm TeV}$ calculated 
 with MCFM~\cite{MCFM} using the PDF sets taken from 
 MSTW2008~\cite{MSTW2008}.
}
\label{TABLE_XS_LHC_Antitop}
\end{table}

Finally, the measurements suggested and outlined here provide 
complementary and assumption-free constraints that can be used 
and combined to those obtained via direct searches of a fourth 
generation and/or precision observables. Work in this direction 
is in progress. 

\section*{ACKNOWLEDGEMENTS}
This work has been performed using the CKMfitter package. 
We would like to thank J. Wagner-Kuhr for useful discussions.
We are grateful for the support provided by the CKMfitter group. 
F. M. would like to thank Jean-Marc G\'erard for many useful
discussions. 
A. M. is funded by the German Science Foundation (DFG). 
F. M. and M. Z. are funded by the Belgian Federal Office for Scientific, 
Technical and Cultural Affairs through Interuniversity Pole No. P6/11.

\end{document}